\documentclass[12pt,a4paper]{article}
\usepackage{xcolor}
\usepackage{jheppub}
\usepackage{epstopdf}
\usepackage{graphicx}
\usepackage{epsfig}
\usepackage{dcolumn}  
\usepackage{bm}    
\usepackage{amssymb} 
\usepackage{amsmath,bm}
\usepackage{amsfonts}    
\usepackage{dsfont}
\usepackage{slashed}  
\usepackage{youngtab}
\usepackage{tikz}
\usepackage{pgf}
\usetikzlibrary{arrows}
\usepackage[mathscr]{euscript}
\usepackage{epsfig}
\usepackage{pdfpages}
\usepackage{verbatim}
\usepackage{multirow}
\usepackage{tensor}

\newdimen\tableauside\tableauside=1.0ex
\newdimen\tableaurule\tableaurule=0.4pt
\newdimen\tableaustep
\def\phantomhrule#1{\hbox{\vbox to0pt{\hrule height\tableaurule width#1\vss}}}
\def\phantomvrule#1{\vbox{\hbox to0pt{\vrule width\tableaurule height#1\hss}}}
\def\sqr{\vbox{%
		\phantomhrule\tableaustep
		\hbox{\phantomvrule\tableaustep\kern\tableaustep\phantomvrule\tableaustep}%
		\hbox{\vbox{\phantomhrule\tableauside}\kern-\tableaurule}}}
\def\squares#1{\hbox{\count0=#1\noindent\loop\sqr
		\advance\count0 by-1 \ifnum\count0>0\repeat}}
\def\tableau#1{\vcenter{\offinterlineskip
		\tableaustep=\tableauside\advance\tableaustep by-\tableaurule
		\kern\normallineskip\hbox
		{\kern\normallineskip\vbox
			{\gettableau#1 0 }%
			\kern\normallineskip\kern\tableaurule}%
		\kern\normallineskip\kern\tableaurule}}
\def\gettableau#1 {\ifnum#1=0\let\next=\null\else
	\squares{#1}\let\next=\gettableau\fi\next}

\allowdisplaybreaks[1]

\newcommand{\be}{ \begin{equation}}
\newcommand{\ee}{\end{equation}}
\newcommand{\bea}[1]{\begin{eqnarray}\label{#1} }
\newcommand{\eea}{\end{eqnarray}}

\def\ZZZ{{\hskip-3pt\hbox{ Z\kern-1.6mm Z}}}
\def\zzz{{\hskip-3pt\hbox{ z\kern-1mm z}}}

\newcommand{\vt}{\vartheta}

\newcommand{\abs}[1]{\left| #1 \right|}

\def\tZ{\hat{Z}}

\newcommand{\half}{{1\over 2}}

\def\one{{\hbox{ 1\kern-.8mm l}}}
\def\zero{{\hbox{ 0\kern-1.5mm 0}}}

\def\vtc#1#2#3{\vartheta _{#1}({#2}|{#3})}
\def\ie{{\it i.e.~}}
\def\eg{{\it e.g.~}}

\def\nn{\nonumber}

\def\l1{{{1-loop}}}

\def\bz{{\bar{z}}}

\def\n1{\Bigg|_{n=1}}

\def\one{{(1)}}
\def\zero{{(0)}}
\def\n{{(n)}}
\def\tr{{Tr}}

\def\cN{\mathcal{N}}

\def\half{\frac{1}{2}}
\def\tr{\text{Tr}}

\def\cZ{\mathcal{Z}}

\def\bL{\bar{L}}

\def\btau{\bar{\tau}}
\def\bq{\bar{q}}

\def\cN{\mathcal{N}}

\def\ie{{\it i.e.~}}
\def\eg{{\it e.g.~}}

\def\nn{\nonumber}

\def\l1{{{1-loop}}}

\def\bz{{\bar{z}}}

\def\n1{\Bigg|_{n=1}}

\def\one{{(1)}}
\def\zero{{(0)}}
\def\n{{(n)}}
\def\tr{{Tr}}

\def\cN{\mathcal{N}}

\def\vt{\vartheta}
\def\half{\frac{1}{2}}
\def\tr{\text{Tr}}

\def\cZ{\mathcal{Z}}

\def\bL{\bar{L}}

\title{Stringy $\mathcal{N}=(2,2)$ holography for AdS${_3}$}

\author{Shouvik Datta,}
\author{Lorenz Eberhardt}
\author{and Matthias R.\ Gaberdiel}
 
\affiliation{Institut f\"ur Theoretische Physik, ETH Zurich, \\
CH-8093 Z\"urich, Switzerland.}
\emailAdd{$\lbrace$shouvik,eberhardtl,gaberdiel$\rbrace$@itp.phys.ethz.ch}

\abstract{We propose a class of ${\rm AdS}_3/{\rm CFT}_2$ dualities with $\mathcal{N}=(2,2)$ supersymmetry. 
These dualities relate string theory on ${\rm AdS}_3 \times ({\rm S}^3\times \mathbb{T}^4)/{\rm G}$ to marginal deformations of the symmetric product orbifold of $\mathbb{T}^4/{\rm G}$, where ${\rm G}$ is a dihedral group. We demonstrate that the BPS spectrum calculated from supergravity and string theory agrees with that of the dual CFT. Moreover, the supergravity elliptic genus is shown to reproduce the CFT answer, thus providing further non-trivial evidence in favour of the proposal. 
}

\makeatletter
\g@addto@macro\bfseries{\boldmath}
\makeatother

\begin{document}

\maketitle

\section{Introduction}

The holographic principle of Maldacena \cite{Maldacena:1997re} has been explored, since its inception, in many different contexts.
One important class of examples are those associated to $\mathrm{AdS}_3$. 
For that case,  there is very good evidence that the CFT dual of string theory on $\mathrm{AdS}_3\times \mathrm{S}^3 \times \mathcal{M}_4$, where $\mathcal{M}_4$ is a hyper-K\"ahler background (\ie K3 or $\mathbb{T}^4$) is the symmetric orbifold of $\mathcal{M}_4$, see, \eg \cite{David:2002wn} for a review. 
More recently, also the CFT dual of string theory on $\mathrm{AdS}_3\times \mathrm{S}^3 \times \mathrm{S}^3 \times \mathrm{S}^1$ has been identified \cite{Eberhardt:2017pty} building on earlier work \cite{Eberhardt:2017fsi,Baggio:2017kza,Elitzur:1998mm,Gukov:2004ym}: at least for certain background charges it is also described by a symmetric product orbifold, namely of $\mathrm{S}^3 \times \mathrm{S}^1$.

All of these examples have (small or large) ${\cal N}=(4,4)$ superconformal symmetry, and for the case of $\mathcal{M}_4 = \mathbb{T}^4$, the relation of the stringy duality to the higher spin/CFT dualities of \cite{Gaberdiel:2010pz} has also been understood \cite{Gaberdiel:2014cha}. 
It is a natural to question whether one can also find examples with lower supersymmetry, in particular ${\cal N}=(2,2)$ superconformal symmetry. 
Among other things, this may allow one to understand how the ${\cal N}=2$ superconformal higher spin/CFT dualities of \cite{Creutzig:2011fe,Candu:2012jq} (or some suitable generalisations of them) may fit into stringy dualities. 
In particular, the Kazama-Suzuki models that appear in these constructions are quite reminiscent of the dual theories of \cite{Chang:2012kt} in one dimension higher, and it would be very interesting to understand how the results of \cite{Chang:2012kt} and \cite{Gaberdiel:2014cha} fit together.

As far as we are aware, no dualities with $\cN=(2,2)$ supersymmetry have been constructed so far. In this work, we find a class of examples which involve specific quotients of ${\rm AdS}_3\times \mathrm{S}^3 \times \mathbb{T}^4$ preserving  $\cN=(2,2)$ supersymmetry. 
The $\cN=(2,2)$ case provides a controllable setting since one can make use of supersymmetry to study protected states and compare quantities on the CFT and gravity sides. Although the dualities we shall propose below do not make direct contact with the above higher-spin/CFT correspondence, the CFTs that appear possess a higher spin symmetry. (This is reminiscent of the situation that was encountered in \cite{Eberhardt:2017pty}.) 

The dualities which we propose in this paper involve string theory on the background 
$
\mathrm{AdS}_3 \times (\mathrm{S}^3 \times \mathbb{T}^4 )/{\rm G}
$,
where ${\rm G} \subset \mathrm{D}_n$, the dihedral group, while the CFTs are symmetric product orbifolds of $ \mathbb{T}^4/{\rm G}$. 
As evidence in favour of these proposals we shall show that the BPS spectrum and the elliptic genus matches. 
We should note that ${\rm G}$ acts also non-trivially on $\mathrm{S}^3 $: in fact, this is required to reduce the supersymmetry from $\cN=(4,4)$ to $\cN=(2,2)$ as 
the R-symmetry needs to be broken from $\mathfrak{su}(2)\oplus \mathfrak{su}(2)$ to $\mathfrak{u}(1)\oplus\mathfrak{u}(1)$. 
As a consequence, various somewhat intriguing subtleties emerge, \eg spacetime supersymmetry is only preserved (and hence the duality only works) provided that the flux through the three-sphere (or rather its quotient) is odd --- this is at least the case for the situation with pure NS-NS flux. Taking a non-abelian orbifold provides us also with the interesting possibility of introducing discrete torsion \cite{Vafa:1986wx}. The duality also works for this modification.

%
\smallskip

The outline of the paper is as follows. In Section \ref{sec:quotients} we provide the details of the string backgrounds which we shall be dealing with. We compute the BPS spectra of the proposed dual CFTs in Section \ref{sec:CFT}, and study their elliptic genus in Section~\ref{sec:elliptic genus}. In Section \ref{sec:sugra}, we confirm that the backgrounds we have constructed support indeed \linebreak $\mathcal{N}=(2,2)$ supersymmetry. Furthermore, we compute the supergravity BPS spectrum. This BPS spectrum is reproduced from a world-sheet string theory analysis (using the WZW model approach to describe AdS$_3$) in Section \ref{sec:string theory}. There we also calculate the contributions from the twisted sectors and compute the supergravity limit of the elliptic genus, which is shown to reproduce the CFT answer. We conclude in Section~\ref{sec:conclusions}. There are a number of appendices where some of the more technical calculations can be found.

\section{Taking quotients of $\mathrm{AdS}_3\times \mathrm{S}^3\times \mathbb{T}^4$}\label{sec:quotients}

In this section we will describe how to obtain $\mathcal{N}=(2,2)$  CFTs from taking a suitable orbifold by a finite group of (the symmetric orbifold of) $\mathbb{T}^4$. We then identify the corresponding orbifold action in the dual string background on $\mathrm{AdS}_3 \times \mathrm{S}^3 \times \mathbb{T}^4$. 

\subsection{The example of $(\mathbb{T}^2/\mathbb{Z}_2)\times \mathbb{T}^2$}\label{simpex}

Recall that one can realize K3 as the orbifold $\mathbb{T}^4/\mathbb{Z}_2$, where the $\mathbb{Z}_2$ acts by inversion of all four coordinates.
There are other finite quotients we can take of the torus $\mathbb{T}^4$ which preserve an $\mathcal{N}=2$ structure, but not the complete $\mathcal{N}=4$ structure --- we will describe a family of them more systematically below. 
A simple example is the theory $(\mathbb{T}^2/\mathbb{Z}_2) \times \mathbb{T}^2$, where the $\mathbb{Z}_2$ action inverts only the first two coordinates. 
Since every $\mathbb{T}^2$ supports separately an $\mathcal{N}=2$ algebra whose generators are constructed entirely out of bilinears of the fundamental fields, the $\mathcal{N}=2$ algebra survives the orbifold projection. 
On the other hand, this orbifold does not preserve the $\mathcal{N}=4$ algebra since, for example, the bilinear fermion terms, involving one fermion from each $\mathbb{T}^2$, are not invariant. (These are the spin $h=1$ fields that extend $\mathfrak{u}(1)$ to $\mathfrak{su}(2)$.)
In analogy to K3, one might naively think that the symmetric orbifold of this CFT should be dual to the background $\mathrm{AdS}_3 \times \mathrm{S}^3 \times (\mathbb{T}^2/\mathbb{Z}_2) \times \mathbb{T}^2$, but this is wrong for a number of reasons. 
First of all, the factor $\mathbb{T}^2/\mathbb{Z}_2$\footnote{This is also known as the `pillow' or `ravioli' in the supergravity literature.} breaks all supersymmetry; indeed, by calculating the supergravity particle spectrum, one sees that there is a different number of bosons and fermions. 
Second, by the usual lore of the AdS/CFT-correspondence, every symmetry of the bulk must correspond to a symmetry in the dual CFT.
For the case of AdS$_3$, the symmetry gets enhanced to an affine symmetry \`a la Brown-Henneaux \cite{Brown:1986nw}, and the isometry group of ${\rm S}^3$ $\mathfrak{so}(4) \cong \mathfrak{su}(2) \oplus \mathfrak{su}(2)$ will give rise to an $\mathfrak{su}(2) \oplus \mathfrak{su}(2)$ affine symmetry in the dual CFT --- but as we have just seen, the spin $h=1$ fields that extend $\mathfrak{u}(1)$ to $\mathfrak{su}(2)$ do not survive in the dual CFT, and indeed the presence of an unbroken $\mathfrak{su}(2)$ symmetry essentially always leads to a background with ${\cal N}=4$ supersymmetry \cite{Banks:1988yz}.

The second problem points towards taking also some orbifold of the three-sphere $\mathrm{S}^3$ in the bulk. 
In fact, given that the roots of $\mathfrak{su}(2)$ in the dual CFT are odd under the $\mathbb{Z}_2$, we should look for an action on the three-sphere that behaves the same way on the isometry generators. 
It is relatively easy to see that the relevant action we are looking for is a rotation by 180 degrees. 
(In fact, there are only two possible $\mathbb{Z}_2$ isometries  on $\mathrm{S}^3$: inversion which yields $\mathrm{SO}(3)$, and rotation by 180 degrees; inspection shows that the latter is the relevant construction.)


Thus we are led to conclude that we should take the same orbifold of the bulk geometry not only on the fundamental fields, but also on the $\mathrm{S}^3$, and hence that the correct dual bulk geometry is 
\be 
\mathrm{AdS}_3 \times( \mathrm{S}^3 \times \mathbb{T}^2)/\mathbb{Z}_2 \times \mathbb{T}_2\ ,
\ee
where the $\mathbb{Z}_2$-action inverts the coordinates of the two-torus $\mathbb{T}^2$, while at the same time rotating the three-sphere $\mathrm{S}^3$ by 180 degrees. This modification resolves both problems we encountered above: the background is $\mathcal{N}=(2,2)$ supersymmetric --- this will be shown in Section~\ref{sec:sugra} ---  and it has the correct isometry group.  

\subsection{The hyperelliptic case} \label{subsec:hyperelliptic surface}

We should mention that this background, as well as the target space of the dual CFT is singular. 
(This is to be contrasted with the case of K3, where the orbifold has a free resolution.) There are various ways to see this. The would-be resolution of $\mathbb{T}^2/\mathbb{Z}_2$ has to be again a smooth Riemann surface. In fact, vanishing of the axial anomaly requires it, as usual, to be Calabi-Yau, but this means it has to be $\mathbb{T}^2$ again. However, we will determine below the BPS spectrum of $\mathbb{T}^2/\mathbb{Z}_2$, and we will see that it differs from that of the torus $\mathbb{T}^2$.  

We can, however, give a modification of our proposal to obtain a duality between a smooth $\sigma$-model and a smooth background. For this, we include a shift in the remaining $\mathbb{T}^2$, which resolves the singularities. The result is called a hyperelliptic surface.\footnote{This is a complex surface and should not be confused with a hyperelliptic Riemann surface.} Thus, we conjecture that the symmetric orbifold of the hyperelliptic surface $(\mathbb{T}^2\times \mathbb{T}^2)/\mathbb{Z}_2$ is dual to string theory on the background geometry
\be 
\mathrm{AdS}_3 \times (\mathrm{S}^3 \times \mathbb{T}^2 \times \mathbb{T}^2)/\mathbb{Z}_2 \ ,
\ee 
where the $\mathbb{Z}_2$ acts now by inverting one $\mathbb{T}^2$ and shifting the other. (On the bulk side, it also rotates the three-sphere by 180 degrees.)
%
%
Because of the quotient action on $\mathbb{T}^2 \times \mathbb{T}^2$, the quotient space develops a torsion cohomology group $\mathrm{H}^2(X;\mathbb{Z})=\mathbb{Z}^2 \oplus \mathbb{Z}_2$, and the first Chern class is precisely the torsion element of the cohomology group. Thus, even though the first Chern class of this K\"ahler manifold is not zero, it vanishes in real cohomology. This still suffices for the space to be Ricci flat (in fact the induced metric is obviously Ricci flat), so there is no axial anomaly. 


To our knowledge, this is the first known supergravity background which supports $\mathcal{N}=(2,2)$ superconformal symmetry. We shall show below that the BPS spectrum of this background matches with that of the dual CFT. (The elliptic genus vanishes on both sides because of the presence of the $\mathbb{T}^2$ factor.)

\subsection{Generalizing to the dihedral group}\label{sec:dihedral}
We can generalize the singular construction of Section~\ref{simpex} by considering the quotient\footnote{In the general case, there is no obvious analogue of the smooth construction of Section~\ref{subsec:hyperelliptic surface}. However, the backgrounds still preserve ${\cal N}=(2,2)$ supersymmetry \cite{Eberhardt:2017uup}.}
\be 
\mathbb{T}^4/{\rm G} \ ,
\ee 
where ${\rm G} \subset \mathrm{D}_n$ is a finite subgroup of the dihedral group; conventions and useful properties of the dihedral group are summarized in Appendix~\ref{app:branching rules}. In the following we shall need that the dihedral group can be generated by two elements, $S$ and $R$ satisfying
\be 
R^n=S^2=(RS)^2=1\ .
\ee
Since $S$ squares to one, we will refer to it as the reflection, while we call $R$ the rotation. Here, $\mathrm{D}_n$ acts in twice the fundamental representation $2\cdot \rho_1$ on $\mathbb{T}^4$. In the fundamental representation of the dihedral group the reflection generator $S$ acts as a reflection in 2-dimensional space, while the rotation $R$ generator describes a rotation with angle $\tfrac{2\pi}{n}$, see  \eqref{rho i} for the precise definition. The lattice of the torus has to be preserved by the action, so some crystallographic constraint arises in the construction, \ie the only possibilities are $n=1$, $2$, $3$, $4$ and $6$, see {\it e.g.}~\cite[Section 5.2]{Grove:1996}. 
Moreover, as may be familiar from two-dimensional point groups, there are two possibilities for the representation of the inversion $S$, which are not conjugate in $\mathrm{GL}(2,\mathbb{Z})$: $S$ can act as either of
\be 
\begin{pmatrix}
-1 & 0 \\ 0 & 1
\end{pmatrix}\ , \qquad 
\begin{pmatrix}
0 & 1 \\ 1 & 0
\end{pmatrix}\ . \label{possibilities for S}
\ee
For $\mathrm{D}_1$, $\mathrm{D}_2$ and $\mathrm{D}_3$ the corresponding actions are non-conjugate in $\mathrm{GL}(2,\mathbb{Z})$, while for $\mathrm{D}_4$ and $\mathrm{D}_6$ they are. As a consequence, in the former case they are not related by a change of basis of the lattice to one another, and hence define  two inequivalent actions.\footnote{Of course, the two actions are equivalent as actions on $\mathbb{R}^4$. In more mathematical terms, we are interested in representations over $\mathbb{Z}$, not in representations over $\mathbb{R}$.} For $i=1, 2, 3$ we will write $\mathrm{D}_{i}^{(1)}$ for the first possibility, and $\mathrm{D}_{i}^{(2)}$ for the second one.
The example of Section~\ref{simpex} corresponds then to the case $\mathrm{D}_{1}^{(1)}$. The explicit matrix realizations of these representations are spelled out in Appendix~\ref{app:integers}.

Of course, when ${\rm G}=\mathbb{Z}_n \subset \mathrm{D}_n$, these are just different incarnations of K3, and they will all lie on the same moduli space. So interesting new examples will only arise by taking ${\rm G}$ to be the full dihedral group. As will become apparent from our results below, at least most of the resulting CFTs are distinct (since their BPS spectra differ).

The main claim of the paper is that the symmetric product orbifold of these theories is dual to the bulk geometry
\be 
\mathrm{AdS}_3 \times (\mathrm{S}^3 \times \mathbb{T}^4)/{\rm G}\ ,
\ee
where ${\rm G}=\mathrm{D}_n$, and the inversion generators of $\mathrm{D}_n$ act on $\mathrm{S}^3$ by a 180 degree rotation.

\subsection{Discrete torsion}
There exists a simple further generalization of the models we have presented so far. Some of them admit discrete torsion, \ie we can consider the partition function
\be 
Z=\frac{1}{|{\rm G}|}\sum_{gh=hg} \epsilon(g,h)\ \text{\scalebox{.8}{$g$}}\underset{h}{\raisebox{-5pt}{\text{\scalebox{2}{$\square$}}}}\ ,
\ee
where $\epsilon(g,h)$ are some non-trivial phases. Modular invariance and factorization at genus two imposes strong constraints on the phases which can appear \cite{Vafa:1986wx}. Non-trivial discrete torsion is classified by the second group cohomology $\mathrm{H}^2({\rm G};\mathrm{U}(1))$ of the finite group ${\rm G}$ we are considering. For the cyclic and dihedral groups one has \cite{Handel:1993pr}
\be 
\mathrm{H}^2(\mathbb{Z}_n;\mathrm{U}(1))=0\ , \quad \mathrm{H}^2(\mathrm{D}_n;\mathrm{U}(1))=\mathbb{Z}_{\mathrm{gcd}(n,2)}\ ,
\ee
\ie the even dihedral groups admit discrete torsion. The case $\mathrm{D}_2 \cong \mathbb{Z}_2 \times \mathbb{Z}_2$ (which defines an abelian orbifold) is  probably best known, and is, for example, discussed in detail in \cite{Gaberdiel:2000fe}.  An explicit formula for $\epsilon(g,h)$ is then given by
\be 
\epsilon(g,h)=\begin{cases}
1 &  g=1 \ \hbox{or} \ h=1  \ \hbox{or} \ g=h  \ \hbox{or} \ (g=R^a \ \hbox{and} \ h=R^b)  \\
-1 & \text{otherwise.}
\end{cases}
\ee
We can modify the above dualities by introducing, both for the bulk string theory as well as for the dual CFT, these discrete torsion phases.\footnote{While a world-sheet description is not directly available, except at the pure NS-NS point where one can describe the background in terms of WZW models, see also Section~\ref{sec:string theory}, it should be possible to introduce discrete torsion phases into any orbifold of a world-sheet CFT.}
It will follow from the subsequent discussion that the matching of the BPS spectrum and the elliptic genus will work equally well for these cases.


\section{BPS spectrum of the CFT} \label{sec:CFT}

In the rest of the paper we will subject these proposals to various consistency checks. In particular, we will compare the BPS spectrum and the elliptic genus. In this section we shall analyze the BPS spectrum of the dual CFTs. The calculation of the elliptic genus will be performed in the following section, while the comparison with supergravity and string theory will be done in Sections~\ref{sec:sugra} and \ref{sec:string theory}, respectively. 

\subsection{BPS spectrum of seed theory without discrete torsion}

In the following, we shall discuss the different examples in turn. For the untwisted sector (with respect to the $\mathrm{D}_n$ action) we can give a fairly uniform treatment, while the contributions of the twisted sectors have to be treated  case by case. 
As we have explained in Section~\ref{sec:dihedral}, the four torus directions of $\mathbb{T}^4$ transform in the $2\cdot \rho_1$ representation of $\mathrm{D}_n$, see Appendix~\ref{app:dihedral groups} for more details about the representation theory of $\mathrm{D}_n$. Thus the different contributions to the Hodge diamond of $\mathbb{T}^4$ transform as 
\be 
\begin{tabular}{ccccccc}
& & & $\rho_+$ & & & \\
& & $\rho_1$ & &$\rho_1$ & & \\
$\rho_-$ & & & $\rho_2 \oplus \rho_- \oplus \rho_+$ & & & $\rho_-$ \\
& & $\rho_1$ & &$\rho_1$ & & \\
& & & $\rho_+$ & & &
\end{tabular}\ . \label{quotient_untwisted_Hodge_diamond}
\ee
For small values of $n$, some of these representations are reducible (or equivalent); in particular, we have for 
\begin{eqnarray}
n=1: & \qquad & \rho_1 \cong \rho_2 \cong \rho_+ \oplus \rho_- \ ,  \label{n=1special}\\
n=2: & \qquad & \rho_2 \cong \rho_+ \oplus \rho_- \ , \label{n=2special}
\end{eqnarray}
while for $n=3$, $\rho_2 \cong \rho_1$. In the untwisted sector then only the singlet states (\ie the states transforming in $\rho_+$) survive. 

For the twisted sector also the action of the reflection generator plays a role, \ie we need to distinguish between $\mathrm{D}_n^{(1)}$ and $\mathrm{D}_n^{(2)}$. (As a consequence, it is not easy to give a very uniform treatment for all cases.) In addition, as we shall see, the twisted sector states will  contribute also to half-integer Hodge numbers --- as we have mentioned before, these CFTs do not give rise  to a supersymmetric background in spacetime by themselves, see the discussion in Section~\ref{simpex}, and this is reflected here in the emergence of half-integer Hodge numbers. Explicitly, the BPS spectra of the different cases turn out to be as follows. 

\subsubsection*{$\mathrm{D}_1^{(1)}$}
This case is simply $(\mathbb{T}^2/\mathbb{Z}_2) \times \mathbb{T}^2$. The BPS spectrum of $\mathbb{T}^2/\mathbb{Z}_2$ is $h_{0,0}=h_{1,1}=1$, $h_{1/2,1/2}=4$ since there are four fixed points of the action. This is analogous to the familiar $\mathbb{Z}_2$-orbifold of $\mathbb{T}^4$, which yields one realization of K3, where we had 16 fixed points. Thus, upon tensoring with $\mathbb{T}^2$, we find the result given in Table~\ref{tab:torus quotients BPS spectra}. Note that, because of  (\ref{n=1special}), the states corresponding to $h_{1,0}=h_{0,1}=h_{2,1}=h_{1,2}=h_{2,2}=1$ all arise from the untwisted sector, as do the two states at $h_{1,1}$. This can also be confirmed independently by doing a character analysis.
\subsubsection*{$\mathrm{D}_1^{(2)}$}
This is essentially the same as the previous case, but now there is only one fixed point, which yields the result of Table~\ref{tab:torus quotients BPS spectra}. The fixed point is the diagonal torus $\Delta(\mathbb{T}^2) \subset \mathbb{T}^2 \times \mathbb{T}^2 \cong \mathbb{T}^4$.
\subsubsection*{$\mathrm{D}_2^{(1)}$}
This is simply $(\mathbb{T}^2/\mathbb{Z}_2) \times (\mathbb{T}^2/\mathbb{Z}_2)$, thus we just have to tensor the BPS spectrum of $\mathbb{T}^2/\mathbb{Z}_2$ with itself, leading to the result of Table~\ref{tab:torus quotients BPS spectra}.

\begin{table}
\begin{center}
\begin{tabular}{cc}\hline\vspace{-.5cm} \\ 
Orbifold action & BPS spectrum \\ \hline
$\mathrm{D}_1^{(1)}$ & \begin{tabular}{ccc}
& 1 & \\
1 & 4 & 1 \\
4 & 2 & 4 \\
1 & 4 & 1 \\
& 1 &
\end{tabular} \\ 
\hline
$\mathrm{D}_1^{(2)}$ & \begin{tabular}{ccc}
& 1 & \\
1 & 1 & 1 \\
1 & 2 & 1 \\
1 & 1 & 1 \\
& 1 &
\end{tabular} \\ 
\hline
$\mathrm{D}_2^{(1)}$ & \begin{tabular}{ccc}
& 1 & \\
0 & 8 & 0 \\
0 & 18 & 0 \\
0 & 8 & 0 \\
& 1 &
\end{tabular} \\ 
\hline
$\mathrm{D}_2^{(2)}$ & \begin{tabular}{ccc}
& 1 & \\
0 & 2 & 0 \\
0 & 12 & 0 \\
0 & 2 & 0 \\
& 1 &
\end{tabular} \\ \hline 
\end{tabular}
\begin{tabular}{cc}\hline\vspace{-.5cm} \\ 
Orbifold action & BPS spectrum \\ \hline 
$\mathrm{D}_3^{(1)}$ & \begin{tabular}{ccc}
& 1 & \\
0 & 1 & 0 \\
1 & 10 & 1 \\
0 & 1 & 0 \\
& 1 & 
\end{tabular} \\ \hline
$\mathrm{D}_3^{(2)}$ & \begin{tabular}{ccc}
& 1 & \\
0 & 1 & 0 \\
1 & 10 & 1 \\
0 & 1 & 0 \\
& 1 &
\end{tabular} \\ \hline
$\mathrm{D}_4$ & \begin{tabular}{ccc}
& 1 & \\
0 & 5 & 0 \\
0 & 15 & 0 \\
0 & 5 & 0 \\
& 1 &
\end{tabular} \\ \hline
$\mathrm{D}_6$ & \begin{tabular}{ccc}
& 1 & \\
0 & 2 & 0 \\
0 & 12 & 0 \\
0 & 2 & 0 \\
& 1 &
\end{tabular} \\  \hline 
\end{tabular} \\
\end{center}
\caption{BPS spectra of the torus orbifolds. The different entries are the non-vanishing Hodge numbers $h_{p,q}$, where $p,q\in\{0,\frac{1}{2},1,\frac{3}{2},2\}$ with $p+q\in\mathbb{N}_0$. Recall that $\mathrm{D}_i$ specifies the orbifold group, which acts in the real represention $2 \cdot \rho_1$. The superscript $(1)$ or $(2)$ determines the $\mathbb{Z}$-representation, as explained around \eqref{possibilities for S}.} \label{tab:torus quotients BPS spectra}
\end{table}

\subsubsection*{$\mathrm{D}_2^{(2)}$}
This is the first slightly more difficult case. We will always denote the generators of $\mathrm{D}_n$ by $R$ and $S$, where $R$ is the rotation generator and $S$ the reflection generator. For $n=2$, the group is still commutative. The contribution of the untwisted sector is $h_{0,0}=h_{2,2}=1$, $h_{1,1}=2$, as follows from (\ref{quotient_untwisted_Hodge_diamond}) with (\ref{n=2special}). Let us consider the $R$ twisted sector. There are again 16 BPS states, since $R$ has 16 fixed points. However, only $4+\tfrac{1}{2}\times 12=10$ of them are also invariant under $S$, so this twisted sector contributes $h_{1,1}=10$. Indeed, the 16 fixed points are (for simplicity we are considering an orthogonal lattice)
\be 
\tfrac{1}{2}(\epsilon_1+\mathrm{i}\epsilon_2 ,\epsilon_3+\mathrm{i}\epsilon_4)\ , \quad \epsilon_i \in \{0,\, 1\}\ ,
\ee 
and $S$ interchanges the first complex coordinate with the second one \eqref{possibilities for S}. Thus the fixed points are
\be 
 \tfrac{1}{2}(\epsilon_1+\mathrm{i}\epsilon_2,\epsilon_1+\mathrm{i}\epsilon_2)\ , \quad \tfrac{1}{2}(\epsilon_1+\mathrm{i}\epsilon_2,\epsilon_3+\mathrm{i}\epsilon_4)+\tfrac{1}{2}(\epsilon_3+\mathrm{i}\epsilon_4,\epsilon_1+\mathrm{i}\epsilon_2)\ .
\ee
The first set contributes 4 fixed points under $S$, while the second set gives rise to $\tfrac{1}{2}\times 12=6$. (Here the linear combination in the second set is to be thought of as a superposition of states in the Hilbert space.)

 Moving on to the $S$ twisted sector, it has the diagonal torus $\Delta(\mathbb{T}^2)\subset \mathbb{T}^2 \times \mathbb{T}^2\cong\mathbb{T}^4$ as a fixed point. $R$ performs a further $\mathbb{Z}_2$ orbifold of this torus, so this sector contributes $h_{1/2,1/2}=h_{3/2,3/2}=1$. Finally, the $RS$ twisted sector has the anti-diagonal torus $\widetilde{\Delta}(\mathbb{T}^2)=\{(z,-z)\} \subset\mathbb{T}^4$ as a fixed point, and $R$ acts again as a $\mathbb{Z}_2$ orbifold on this torus. So the $RS$ twisted sector contributes the same amount as the $S$ twisted sector. In total we obtain the result given in Table~\ref{tab:torus quotients BPS spectra}.
\medskip

Following \cite[Section 17.B.4]{DiFrancesco:1997nk}, all other cases can be reduced to these. To see this more explicitly, let us first consider the case of a dihedral group $\mathrm{D}_n$ of odd order ($n$ odd). Then one finds for the partition function
\be 
Z_{\mathrm{D}_n}=\frac{1}{2}(Z_{\mathbb{Z}_n}+2Z_{\mathrm{D}_1}-Z)\ . \label{dihedral group odd}
\ee
Thus, we obtain immediately the result for $\mathrm{D}_3^{(1)}$ and $\mathrm{D}_3^{(2)}$. By inspection of the involved matrices, we see that all $\mathrm{D}_1$ appearing in the partition function are of type (2), see Appendix~\ref{app:integers}. Hence, there is no distinction between $\mathrm{D}_3^{(1)}$ and $\mathrm{D}_3^{(2)}$ on the level of the BPS spectrum. 

For the other case, namely when the dihedral group has even order, we find instead of (\ref{dihedral group odd}) 
\be 
Z_{\mathrm{D}_n}=\frac{1}{2n}(nZ_{\mathbb{Z}_n}+\tfrac{n}{2}(4 Z_{\mathrm{D}_2}-2Z_{\mathbb{Z}_2}))\sim\frac{2}{n} \frac{n}{2} Z_{\mathrm{D}_2}\ , \label{dihedral group even}
\ee
where the last equality is understood to hold only on the level of the BPS spectrum, since all $\mathbb{Z}_n$ for $n \ge 2$ describe $\mathrm{K}3$ surfaces (and hence $nZ_{\mathbb{Z}_n} - \tfrac{n}{2} 2Z_{\mathbb{Z}_2}\sim 0$). 

One can check (by doing a case by case analysis, see again Appendix~\ref{app:integers}) that for $\mathrm{D}_4$, one $\mathrm{D}_2$ is of type (1), while the other is of type (2). On the other hand, for $\mathrm{D}_6$ all of them are of type (2). 
The formulas \eqref{dihedral group odd} and \eqref{dihedral group even} give then the remaining cases of Table~\ref{tab:torus quotients BPS spectra}.

\subsection{BPS spectrum of torus quotients with discrete torsion}
We can repeat the above analysis for dihedral groups  with the inclusion of discrete torsion (for even $n$), and the results are summarized in Table~\ref{tab:torus quotients BPS spectra discrete torsion}. The dihedral group including discrete torsion will in the following be denoted by $\widetilde{\mathrm{D}}_n$.

\begin{table}[t]
\begin{center}
\begin{tabular}{cc} \hline\vspace{-.5cm} \\ 
Orbifold action & BPS spectrum \\ \hline
$\widetilde{\mathrm{D}}_2^{(1)}$ & \begin{tabular}{ccc}
& 1 & \\
0 & 0 & 0 \\
8 & 2 & 8 \\
0 & 0 & 0 \\
& 1 &
\end{tabular} \\ \hline
$\widetilde{\mathrm{D}}_2^{(2)}$ & \begin{tabular}{ccc}
& 1 & \\
0 & 0 & 0 \\
2 & 8 & 2 \\
0 & 0 & 0 \\
& 1 &
\end{tabular} \\ \hline 
\end{tabular}
\begin{tabular}{cc} \hline\vspace{-.5cm} \\ 
Orbifold action & BPS spectrum \\ \hline 
$\widetilde{\mathrm{D}}_4$ & \begin{tabular}{ccc}
& 1 & \\
0 & 0 & 0 \\
5 & 5 & 5 \\
0 & 0 & 0 \\
& 1 &
\end{tabular} \\ \hline
$\widetilde{\mathrm{D}}_6$ & \begin{tabular}{ccc}
& 1 & \\
0 & 0 & 0 \\
2 & 8 & 2 \\
0 & 0 & 0 \\
& 1 &
\end{tabular} \\ \hline 
\end{tabular}
\end{center}
\caption{BPS spectra of the torus orbifolds with the inclusion of discrete torsion. Again, the entries are the non-vanishing Hodge numbers $h_{p,q}$, where $p,q\in\{0,\frac{1}{2},1,\frac{3}{2},2\}$ with $p+q\in\mathbb{N}_0$.} \label{tab:torus quotients BPS spectra discrete torsion}
\end{table}

%

\subsection{The symmetric orbifold}

So far we have only described the BPS spectrum of the seed theory. Now we want to put these results together to describe the BPS spectrum of the associated symmetric product orbifold. It follows from the DMVV formula \cite{Dijkgraaf:1996xw} that the single-particle BPS spectrum of the symmetric product orbifold is given by overlying the Hodge diamonds on top of each other. For example, for the case of $\mathrm{D}_2^{(1)}$ that we shall sometimes concentrate on, the single-particle BPS spectrum of the symmetric product orbifold has then the Hodge numbers
\be
h_{0,0} =1 \ , \quad  h_{1/2,1/2} = 8 \ , \quad h_{1,1} = 19 \ , \quad h_{n,n} = \left\{
\begin{array}{cl}
16 & \ \ \hbox{$n\geq \frac{3}{2}$ and $n$ half-integer} \\
20 & \ \ \hbox{$n\geq 2$ and $n$ integer.} 
\end{array}
\right.
\ee
The other cases can be calculated similarly. 

Given that we have a uniform description of the untwisted sector, see eq.~(\ref{quotient_untwisted_Hodge_diamond}) above, we can also give a uniform description of the untwisted sector contribution to the whole symmetric orbifold; this leads to 
\begin{align} 
h_{0,0}&=\rho_+\ , \quad h_{1,0}=h_{0,1}=\rho_1\ , \quad h_{1,1}=2\rho_+ \oplus \rho_- \oplus \rho_2\ , \quad h_{n,n-2}=\rho_-\text{ for $n \ge 2$}\ , \nonumber\\
h_{n,n-1}&=h_{n-1,n}=2\rho_1\text{ for $n \ge 2$}\ , \quad h_{n,n}=3\rho_+ \oplus \rho_- \oplus \rho_2\text{ for $n \ge 2$}\ . \label{symmetric orbifold untwisted BPS spectrum}
\end{align}

\section{Elliptic genus of the CFT}\label{sec:elliptic genus}

The next step is to calculate the elliptic genus of the corresponding symmetric orbifolds. We begin by calculating the elliptic genus for the seed theories.

\subsection{Elliptic genus}\label{sec:ell}

As we have mentioned before, there are non-trivial Hodge numbers $h_{p,q}$ for non-integer $(p,q)$. From a CFT viewpoint this means that the $\mathrm{U}(1)$ charges are not all integers. 
As a consequence, the elliptic genus $Z(z,\tau)$ will not define a weak Jacobi form. Recall that a function $\phi(z|\tau)$ defines a weak Jacobi form of weight $w$ and index $m$, if it satisfies
\begin{align}
 \phi\Bigl({z \over  c \tau + d}\, \Bigl| \, {a \tau + b \over  c \tau + d} \Bigr) &=
(c \tau+d)^w e^{ 2 \pi \mathrm{i} m { c z^2 \over  c \tau + d} } \, \phi(z\, |\, \tau)
\qquad \begin{pmatrix} a & b \\ c & d \\ \end{pmatrix} \in \mathrm{SL}(2,\mathbb{Z}) \ ,\label{eq:jactmn1} \\
 \phi(z+ \lambda \tau + \mu \, | \, \tau) &= e^{-2 \pi \mathrm{i} m                                                                                            
(\lambda^2 \tau+ 2 \lambda z)} \, \phi(z\, |\, \tau) \qquad\qquad\qquad
\lambda,\, \mu \in \mathbb{Z} \ , \label{eq:jactmn2}
\end{align}
and has a Fourier expansion
\begin{equation}\label{Fourier}
 \phi(z\, |\, \tau) = \sum_{n \geq 0,\, \ell\in \mathbb{Z}} c(n,\ell) \, q^n y^\ell 
\end{equation}
with $c(n,\ell) = (-1)^w c(n,-\ell)$. (Here $q=e^{2\pi \mathrm{i}\tau}$ and $y=e^{2\pi \mathrm{i} z}$.) The elliptic genus associated to the spaces we are interested in contains also Fourier modes with $\ell\in \frac{1}{2}\mathbb{Z}$ in the expansion (\ref{Fourier}). As a consequence it also does not satisfy the translation property (\ref{eq:jactmn2}), \ie it is not invariant under the shift corresponding to $\mu\in \mathbb{Z}$, \linebreak but only under $\mu \in 2\mathbb{Z}$.
On the other hand, it satisfies the other properties (with $w=0$ and $m=1$) --- this just follows by usual conformal field theory considerations, following essentially the old arguments of \cite{Kawai:1993jk}. In order to be able to use the powerful machinery of weak Jacobi forms, see in particular \cite{zagier},  we define 
\be
\hat{Z}(z, \tau) = Z(2z,\tau)\ .
\ee
By construction, $\hat{Z}(z,\tau)$ then satisfies (\ref{Fourier}), and one easily confirms that also (\ref{eq:jactmn1}) and (\ref{eq:jactmn2}) hold, except that $\hat{m}=4m=4$. Thus the doubled elliptic genus {\em is} a weak Jacobi form, except that it is now at weight $w=0$ and index $m=4$. Then we can use the theory of weak Jacobi forms \cite{zagier}, in particular, the fact that the ring of weak Jacobi forms (of even weight) is freely generated by the Eisenstein series $E_4$, $E_6$ and two forms $\phi_{1,0}$ and $\phi_{1,-2}$ (with index $m=1$ and weight $w=0$ and $w=-2$, respectively), see \cite{zagier} for more details. Thus, the space of weak Jacobi forms of index $m=4$ and weight $w=0$ is four-dimensional, and we can choose a basis as 
\begin{align}
\psi_1(z|\tau)&= \left[\ \sum_{i=2,\, 3,\, 4}\left(\frac{\vartheta_i(z|\tau)}{\vartheta_i(\tau)}\right)^2 \right]^4\ , \\
\psi_2(z|\tau)&= \left[\ \sum_{i=2,\, 3,\, 4}\left(\frac{\vartheta_i(z|\tau)}{\vartheta_i(\tau)}\right)^4 \right]^2\ , \\
\psi_3(z|\tau)&=\sum_{i=2,\, 3,\, 4}\left(\frac{\vartheta_i(z|\tau)}{\vartheta_i(\tau)}\right)^8 \ , \\
\psi_4(z|\tau)&= \left[\ \sum_{i=2,\, 3,\, 4}\left(\frac{\vartheta_i(z|\tau)}{\vartheta_i(\tau)}\right)^2 \right]^2 \left[\ \sum_{i=2,\, 3,\, 4}\left(\frac{\vartheta_i(z|\tau)}{\vartheta_i(\tau)}\right)^4 \right] \ . 
\end{align}
The doubled elliptic genus then has the form
\be 
\hat{Z}(z,\tau)=a\, \psi_1(z|\tau)+b\, \psi_2(z|\tau)+c\, \psi_3(z|\tau)+d \,\psi_4(z|\tau)\ . \label{doubled elliptic genus}
\ee
We now look at the $q^0$ term of $\hat{Z}(z,\tau)$, which contains the information about the BPS states. It is given by
\begin{align}
\hat{Z}(z,\tau)&=\frac{1}{256}(a+b+c+d)\left(y^4+y^{-4}\right)+\frac{1}{32}(5a+b+c+3d)\left(y^3+y^{-3}\right)\nonumber\\
&\qquad+\frac{1}{64}(151a+23b+7c+55d)\left(y^2+y^{-2}\right)\nonumber\\
&\qquad+\frac{1}{32}(515a+39b+7c+149d)\left(y+y^{-1}\right)\nonumber\\
&\qquad+\frac{1}{128}(5603a+739b+291c+2019d)+\mathcal{O}(q)\ .
\end{align}
Since there are no BPS states with charges $y^{2}$ or $y^{3/2}$, the coefficients of $y^{\pm 4}$ and $y^{\pm 3}$ must vanish. Furthermore, the coefficient of $y^{\pm 2}$ is given by $h_{0,0}-h_{0,1}+h_{0,2}$, the coefficient of $y^{\pm 1}$ is given by $h_{1/2,1/2}-h_{1/2,3/2}$, and the constant part is given by $-h_{1,0}+h_{1,1}-h_{1,2}$. Thus, we get an overconstrained system, which leads to a condition on the Hodge numbers 
\be 
8h_{0,1}+h_{1,1}-10h_{0,2}+h_{1/2,3/2}-h_{1/2,1/2}=10\ .
\ee
\begin{table}
\begin{center}
\begin{tabular}{ccccccccccc}\hline\vspace{-.5cm} \\ 
Model & $\mathrm{D}_2^{(1)}$ & $\mathrm{D}_2^{(2)}$ & $\mathrm{D}_3^{(1)}$ & $\mathrm{D}_3^{(2)}$ & $\mathrm{D}_4$ & $\mathrm{D}_6$ & $\widetilde{\mathrm{D}}_2^{(1)}$ & $\widetilde{\mathrm{D}}_2^{(2)}$ & $\widetilde{\mathrm{D}}_4$ & $\widetilde{\mathrm{D}}_6$\\ 
\hline  
$a$ & $1$ & $-\tfrac{1}{2}$ & $-1$ & $-1$ & $\tfrac{1}{4}$ & $-\tfrac{1}{2}$ & $-3$ & $-\tfrac{3}{2}$ & $-\tfrac{9}{4}$ & $-\tfrac{3}{2}$ \\ \hline 
\end{tabular}
\end{center}
\caption{Parameters of the elliptic genera of the different models. The doubled elliptic genus is then given by \eqref{doubled elliptic genus}. The theories with tilde are the versions with discrete torsion.}\label{tab:double elliptic genus}
\end{table}
It is a nice consistency check that this constraint is obeyed by all the BPS spectra we determined above, see Table~\ref{tab:torus quotients BPS spectra} and \ref{tab:torus quotients BPS spectra discrete torsion}. Actually, the relation 
\be 
3a+b=4-4h_{0,1}\overset{n \ge 2 \atop }{=}4
\ee
holds (for $n\geq 2$), and thus, provided that the elliptic genus is non-vanishing (\ie for $n\geq 2$), we can describe it by a single parameter. We choose $a$ to be this independent parameter; the others are then expressed as 
\be 
b=4-3a\ , \quad c=4a-4\ , \quad d=-2a\ . \label{abcd relations}
\ee
The values of the constant $a$ are tabulated in Table~\ref{tab:double elliptic genus}.

\subsection{The specific example of $(\mathbb{T}^2/\mathbb{Z}_2) \times (\mathbb{T}^2/\mathbb{Z}_2)$} 

We can obviously also calculate the elliptic genus of the seed theory directly. We illustrate this and many other things that follow with the example of $\mathrm{D}_2^{(1)}$. Recall that the elliptic genus is defined as the sum over the states in the R-R sector 
\begin{align}
 Z( z,\tau; 0,\btau) = \tr_{\rm RR} \left[ (-1)^F z^{J_0} q^{L_0} \bq^{\bar L_0}  \right] \ .
\end{align}
For  the case at hand, the CFT is a product of the two orbifolds $\mathbb{T}^2/\mathbb{Z}_2$, and we can calculate the elliptic genus separately for the two factors.
For each $\mathbb{T}^2/\mathbb{Z}_2$ we decompose the trace into the contributions coming from the twisted and untwisted sector as 
\begin{align}\label{eg}
 Z(z,\tau;\bz,\btau) = \tr \left[\tfrac{1+{\cal I}}{2}((-1)^F z^{J_0} q^{L_0}  \bq ^{\bL_0})\right]_{\rm U} + 4\, \tr\left[  \tfrac{1+{\cal I}}{2} ((-1)^F z^{J_0} q^{L_0}  \bq ^{\bL_0})\right]_{\rm T} \ ,
\end{align}
where ${\cal I}$ denotes the inversion of the two torus directions, and the factor of $4$ in the contribution of the twisted sector arises from the fact that there are $2^2=4$ fixed points for  $\mathbb{T}^2$ by $\mathbb{Z}_2$. 
In the untwisted sector, the  non-trivial contribution comes from the insertion of ${\cal I}$ for which we find 
\begin{align}
\tr \left[{\cal I}((-1)^F z^{J_0} q^{L_0}  \bq ^{\bL_0})\right]_{\rm U} = 4\frac{\vtc{2}{z}{\tau}}{\vtc{2}{0}{\tau}}\ .
\end{align}
The remaining contributions can be deduced from demanding modular invariance, but we can also calculate them directly. In particular, we find 
\begin{align}
\tr\left[  ((-1)^F z^{J_0} q^{L_0}  \bq ^{\bL_0})\right]_{\rm T}  &= \frac{\vtc{4}{z}{\tau}}{\vtc{4}{0}{\tau}}, \\
\tr\left[{\cal I} ((-1)^F z^{J_0} q^{L_0}  \bq ^{\bL_0})  \right]_{\rm T}  &= \frac{\vtc{3}{z}{\tau}}{\vtc{3}{0}{\tau}}\ .
\end{align}
Because of the half-integer powers of $y$, we need to be careful about which branch we should choose, but the correct (modular invariant) combination turns out to be 
\begin{align}\label{EG-standard}
 {Z}(z,\tau) =   2\left[\frac{\vtc{2}{z}{\tau}}{\vtc{2}{0}{\tau}} + \frac{\vtc{3}{z}{\tau}}{\vtc{3}{0}{\tau}} +
\frac{\vtc{4}{z}{\tau}}{\vtc{4}{0}{\tau}} \right] . 
\end{align}
The elliptic genus of the tensor product $(\mathbb{T}^2 /\mathbb{Z}_2)\times (\mathbb{T}^2 /\mathbb{Z}_2)$ is then simply given by the square of the above expression. The corresponding doubled elliptic genus --- see the discussion in Section~\ref{sec:ell} --- is then indeed a weak Jacobi form, and it is now of index $m=2$, since we are only looking at one copy of $\mathbb{T}^2 /\mathbb{Z}_2$. In fact, one finds explicitly that 
\begin{align}\label{4.20}
Z(z,\tau)=  
2\left[\frac{\vtc{2}{z}{\tau}}{\vtc{2}{0}{\tau}} + \frac{\vtc{3}{z}{\tau}}{\vtc{3}{0}{\tau}} +
\frac{\vtc{4}{z}{\tau}}{\vtc{4}{0}{\tau}} \right] =   \frac{1}{24}\Bigl[  \phi_{0,1}(\tfrac{z}{2},\tau )^2 - \phi_{-2,1}(\tfrac{z}{2},\tau)^2E_4(\tau )\Bigr] \ . 
\end{align}
The square of eq.~(\ref{4.20}), evaluated at $z\mapsto 2z$, agrees then exactly with (\ref{doubled elliptic genus}) for $a=b=1$, $c=0$ and $d=-2$. (In order to see this, one has to use theta function identities, see \eg \cite[Section 8.199]{GR}.)

\subsection{Symmetric orbifold}

Given the elliptic genus of the seed theory, the elliptic genus of the symmetric product orbifold can be read off from the DMVV formula \cite{Dijkgraaf:1996xw}, 
\begin{align}\label{dmvv}
\sum_{N=0}^{\infty} p^N \cZ (X^{\otimes N}/S_N)= \prod_{m=1}^{\infty} \prod_{\Delta,\ell} \frac{1}{\left(1-p^m q^{\Delta/m}y^\ell\right)^{c(\Delta,\ell)}}\ , 
\end{align}
where $c(\Delta,\ell)$ are the coefficients of the elliptic genus of the seed theory $X$,
\begin{align}\label{cexp}
 {Z}_{X}(z,\tau) 
=  \sum_{\Delta,\ell} c(\Delta,\ell)q^{\Delta}y^\ell\ .
\end{align}

Before discussing the general case, let us first consider again the special case where $X = (\mathbb{T}^2 /\mathbb{Z}_2) \times (\mathbb{T}^2 /\mathbb{Z}_2)$. It is convenient to organize the 
elliptic genus according to the sectors from which it originates as 
\begin{align}\label{sep}
 {Z}_{X}(z,\tau)\ =  {Z}_{\mathrm{UU}}(z,\tau) +  {Z}_{\mathrm{TT}}(z,\tau) +  {Z}_{\mathrm{UT}}(z,\tau)\ ,
\end{align}
where
\begin{align}
  {Z}_{\mathrm{UU}}(z,\tau) &= 4 \left[ \frac{\vtc{2}{z}{\tau}}{\vtc{2}{0}{\tau}}  \right] ^2 \ , \qquad  {Z}_{\mathrm{TT}}(z,\tau)  = 4 \left[ \frac{\vtc{3}{z}{\tau}}{\vtc{3}{0}{\tau}} +
\frac{\vtc{4}{z}{\tau}}{\vtc{4}{0}{\tau}} \right] ^2 \ ,
\end{align}
are the contributions where for both $\mathbb{T}^2$'s the contribution is from the untwisted (UU) or twisted (TT) sector, while 
\begin{align}
 {Z}_{\mathrm{UT}}(z,\tau) &= 8\frac{\vtc{2}{z}{\tau}}{\vtc{2}{0}{\tau}} \left[ \frac{\vtc{3}{z}{\tau}}{\vtc{3}{0}{\tau}} +
\frac{\vtc{4}{z}{\tau}}{\vtc{4}{0}{\tau}}  \right] \   
\end{align}
is the mixed contribution (where we have the contribution of the untwisted sector of one $\mathbb{T}^2$, and the twisted one of the other). Given the structure of (\ref{dmvv}), the elliptic genus of the symmetric product is then the product of the contributions of the form (\ref{dmvv}) for each sector separately. 

For each of these sectors, we define the doubled elliptic genus $\hat{Z}_*(z,\tau) = Z_*(2z,\tau)$, where `$*$' stands for 
UU, UT and TT, respectively. From the explicit expressions it is immediate that the expansion coefficients $c_{*}(m,\ell)$ of $\hat{Z}_*(z,\tau)$  (that are defined analogously to (\ref{cexp})) obey the periodicity property, 
\begin{align}\label{period-prop}
c_{*}(m,\ell) = c_*(16m-\ell^2,\ell\text{ mod }8)\ . 
\end{align}
In fact, this just follows from the elliptic translation properties of $\hat{Z}_*(z,\tau)$, \ie from eq.~(\ref{eq:jactmn2}), see \eg \cite[Theorem 2.2]{zagier}  --- this is true sector by sector. (Recall that $\hat{Z}$ has index $m=4$; incidentally, the $m$ in (\ref{period-prop}) is a Fourier mode, and should not be confused with the index.)

In order to compare to supergravity (or string theory), we are interested in the NS-NS sector version of the elliptic genus. This can be obtained from the DMVV formula (\ref{dmvv}) via spectral flow,
\begin{align}\label{dmvvNS}
\sum_{N\geq0} \hat{\cZ}_{\mathrm{NS}} (X^{\otimes N}/S_N)p^N = \prod_{n>0,m,\ell} \frac{1}{(1-p^n q^my^\ell)^{d(n,m,\ell)}} \ . 
\end{align}
The product is over $m,\, \ell$ which obey $m \in \mathbb{Z}_{\geq 0}/4 \, (=\mathbb{Z}_{\geq 0}/2  \,\cup \, (2\mathbb{Z}_{\geq 0}+1)/4) $, the constraint $\ell-4m \in 4 \mathbb{Z}$ and $4m \geq |\ell|$. The $d(n,m,\ell)$ are related to the coefficients of the elliptic genus of the seed theory $c(m,\ell)$ as 
\begin{align}\label{dcrel}
d(n,m,\ell) \ =\  c(n(m-\tfrac{\ell}{4}),\ell-2n)\ =\  c(16mn -\ell^2-4n^2, (\ell-2n)\text{ mod }8) \ . 
\end{align}
Here we have used, in the last identity, eq.~\eqref{period-prop}.
 \begin{table}[!h]
  \begin{center}
  \begin{minipage}{.33\textwidth}
  \begin{center} \rule{1.75in}{.2mm}\\ \vspace{-.2cm}
  UU sector  \\   \vspace{.04cm}
  \begin{tabular}{cc}\hline  
  $\ell=0$ & $1-3\delta_{m,\mathbb{Z}_{>0}}$ \\
  $\abs{\ell}=1$ & 0 \\
  $\abs{\ell}=2$ &  $3-4 \delta_{m,\mathbb{Z}_{>0}+\frac{1}{2}}$ \\
  $\abs{\ell} \ge 3$ & $ 4\delta_{m,\pm l/4}\delta_{m, \frac{1}{2}\mathbb{Z}}$ \\
  \hline 
  \end{tabular}
  \end{center}
  \end{minipage}
  \begin{minipage}{.33\textwidth} \begin{center}\rule{1.88in}{.2mm}\\ \vspace{-.2cm}
    UT sector  \\   \vspace{.04cm}
  \begin{tabular}{cc}\hline  
  $\ell=0$ & $0$ \\
  $\abs{\ell}=1$ & $8-16 \delta_{m-\ell/4,\mathbb{Z}_{>0}}$ \\
  $\abs{\ell} \ge 2$ & $ 16\delta_{m,\ell/4}\delta_{m, \frac{1}{2}\mathbb{Z}+\frac{1}{4}}$ \\
  & \\ \hline 
  \end{tabular}\end{center}
  \end{minipage} 
  \begin{minipage}{.3\textwidth} \begin{center}
  \rule{1.7in}{.2mm}\\ \vspace{-.2cm}
    TT sector  \\   \vspace{.04cm}
  \begin{tabular}{cc}\hline  
  $\ell=0$ & $-16\delta_{m,\mathbb{Z}_{>0}}$ \\
  $\abs{\ell} \ge 1$ & $ 16\delta_{m,\ell/4}\delta_{m, \frac{1}{2}\mathbb{Z}}$\\
  & $\!\!\!\!+16\delta_{m,-\ell/4}\delta_{m,\mathbb{Z}}$  \\
  & \\
  \hline 
  \end{tabular}\end{center}
  \end{minipage}
  \end{center}
  \caption{Values of $\sum_{n>0} d_*(n,m,l)$ appearing in the elliptic genus of the symmetric orbifold of $(\mathbb{T}^2/\mathbb{Z}_2)^2$.} \label{tab:sum-of-coefficients}
  \end{table}
\begin{table}[!b]
  \begin{center}
  \begin{tabular}{cc}\hline  
  $\ell=0$ & $1-(15+4a)\delta_{m,\mathbb{Z}_{>0}}$ \\
  $\abs{\ell}=1$ & $4(a+1)(1-2\delta_{m-\ell/4,\mathbb{Z}_{>0}})$ \\
  ${\ell}=2$ &  $15+4a-(16+4a) \delta_{m,\mathbb{Z}_{>0}+\frac{1}{2}}$ \\
   ${\ell}=-2$ &  $7-4a-(8-4a) \delta_{m,\mathbb{Z}_{>0}+\frac{1}{2}}$ \\
  \multirow{2}{*}{$\abs{l} \ge 3$} & $ (16+4a)\delta_{m,\frac{1}{2}\mathbb{Z}}\delta_{m,\ell/4}+(16+4a)\delta_{m,\mathbb{Z}} \delta_{m,-\ell/4}$\\
  & $+(8-4a)\delta_{m,\mathbb{Z}+\frac{1}{2}}\delta_{m,-\ell/4}+(8a+8)\delta_{m,\ell/4}\delta_{m,\frac{1}{2}\mathbb{Z}+\frac{1}{4}}$ \\
  \hline 
  \end{tabular}
  \end{center}
  \caption{Values of $\sum_{n>0} d(n,m,\ell)$ appearing in the elliptic genus of the symmetric orbifold of $\mathbb{T}^4/{\rm D}_n$.}\label{tab:coefficients general elliptic genus}
\end{table}

To make contact with the supergravity calculations, we shall consider the large $N$ limit, {\it i.e.}\ we consider infinitely many copies of the seed theory, permuted under the infinite symmetric group. The elliptic genus of the symmetric orbifold in the $N\to \infty$ limit can be extracted from the $p\to 1$ limit of \eqref{dmvvNS}, and as shown in \cite{deBoer:1998us}, it leads to 
%
\begin{align}
\hat{\cZ}_{\mathrm{NS}}(z,\tau)= \prod_{(m,\ell) \ne (0,0)} \frac{1}{(1-  q^my^\ell)^{\sum_{n> 0 }d(n,m,\ell)}} \ . 
\end{align}
Note that contrary to the case of K3 \cite{deBoer:1998us}, the elliptic genus does not diverge in the limit $N \to \infty$. (This is a consequence of the fact that in our case $d(1,0,0)=1$, whereas for K3 $d(1,0,0)=2$.)
Thus we simply need to compute the quantity 
\be
\sum_{n>0}d(n,m,\ell) \ . 
\ee 
 Again, we will do this sector by sector. Using eq.~(\ref{dcrel}), we find 
\begin{align}
\sum_{n> 0}d_*(n,m,\ell) &= \sum_{n> 0}  c_*(16m^2 -\ell^2-(2n-4m)^2, (\ell-2n)\text{ mod }8) \nn \\
 &= \sum_{\tilde n>-4m}  c_*(16m^2 -\ell^2- \tilde n^2, (\ell-\tilde{n} -4m)\text{ mod }8) \ ,
\end{align}
where $\tilde{n} = 2n -4m$, and the sum over $\tilde{n}$ runs over even or odd integers (depending on whether $m \in \mathbb{Z}_{\geq 0}/2 $  or 
$ m \in  (2\mathbb{Z}_{\geq 0}+1)/4 $, respectively). 
We can use the property $c_*(m,\ell)=c_*(m,-\ell)$ (which follows from quasiperiodicity in  the elliptic variable $z$, see eq.~(\ref{eq:jactmn2})) to conclude that 
\begin{align}\label{d-sum}
\sum_{n> 0}d_*(n,m,\ell) 
 &= \sum_{\tilde n>-4m}  c_*(16m^2 -\ell^2- \tilde n^2, (\tilde{n}-\ell +4m)\text{ mod }8) \ . 
\end{align}
Next we recall that  $\ell-4m \in 4 \mathbb{Z}$, which implies, $\tilde{n}-\ell+4m\in \lbrace8\mathbb{Z} \cup 8\mathbb{Z}+4\rbrace$. 
We shall consider the cases where $\tilde{n}$ runs over even and odd integers separately --- as mentioned above, this depends on whether $m \in \mathbb{Z}_{\geq 0}/2 $ 
or $ m \in  (2\mathbb{Z}_{\geq 0}+1)/4 $.
Using the results from Appendix~\ref{appC}, see in particular eqs.~\eqref{l mod 8} and \eqref{l+4 mod 8}, we have in the UU sector 
\begin{align}
\sum_{\tilde n\, \in \text{ even }}  c_{\rm UU}(16m^2 -\ell^2- \tilde n^2, (\tilde{n}-\ell +4m)\text{ mod }8) &=  
4 \delta_{m,\pm \ell/4} \quad \text{for }\ell-4m \in 4\mathbb{Z}\ , \nn   \\
\sum_{\tilde n\, \in \text{ odd }}  c_{\rm UU}(16m^2 -\ell^2- \tilde n^2, (\tilde{n}-\ell +4m)\text{ mod }8) &= 
0 \quad \text{for }\ell-4m \in 4\mathbb{Z} \  .
\end{align}
The sums for the diagonally twisted sector are 
\begin{align}
\sum_{\tilde n\, \in \text{ even }}  c_{\rm TT}(16m^2 -\ell^2- \tilde n^2, (\tilde{n}-\ell +4m)\text{ mod }8) &=  \begin{cases}
16 \delta_{m,\pm \ell/4} \quad \text{for }\ell-4m \in 8\mathbb{Z}\ , \nn   \\
0 \quad \, \ \text{for }\ell-4m \in 8\mathbb{Z}+4\  ,\nn   
\end{cases} \\
\sum_{\tilde n\, \in \text{ odd }}  c_{\rm TT}(16m^2 -\ell^2- \tilde n^2, (\tilde{n}-\ell +4m)\text{ mod }8) &= 
0 \quad \text{for }\ell-4m \in 4\mathbb{Z} \ ,
\end{align}
and that of the mixed sector are 
\begin{align}
\sum_{\tilde n\, \in \text{ even }}  c_{\rm UT}(16m^2 -\ell^2- \tilde n^2, (\tilde{n}-\ell +4m)\text{ mod }8) &= 0 \quad \text{for }\ell-4m \in 4\mathbb{Z}\ , \nn \\
\sum_{\tilde n\, \in \text{ odd }}  c_{\rm UT}(16m^2 -\ell^2- \tilde n^2, (\tilde{n}-\ell +4m)\text{ mod }8) &= \begin{cases}
16 \delta_{m,\pm l/4} \quad \text{for }\ell-4m \in 8\mathbb{Z} \  ,  \\
0 \quad \, \ \text{for }\ell-4m \in 8\mathbb{Z}+4 \  .
\end{cases} 
\end{align}
The sum in \eqref{d-sum} starts out at $\tilde n=-4m+2$, whereas the formulas above assume a summation of $\tilde{n}$ over all even or odd integers, respectively. However, for $\abs{\ell}\geq 3$, one can see that all omitted coefficients vanish and we can use the above formulas.\footnote{For small values of $\abs{\ell}$, we need to correct the missing terms by hand; this can be done as in \cite{deBoer:1998us}, see the comments below eq.~(5.6).}
Putting everything together, we then arrive at the final result which is summarized in Fig.~\ref{tab:sum-of-coefficients}. 

The analysis can also be generalized for the other ${\rm D}_n$ orbifolds. 
When orbifolding by $\mathrm{D}_1^{(i)}$, the elliptic genus vanishes, so let us assume $n \ge 2$. Using the relations \eqref{abcd relations}, we can express the answer in terms of the parameter $a$. The analysis precisely parallels the analysis we have done above, and 
the result is given in  Fig.~\ref{tab:coefficients general elliptic genus}.



\section{Supergravity} \label{sec:sugra}

In this section, we analyze these backgrounds from the viewpoint of supergravity. 

\subsection{Killing spinors}

Let us begin by confirming that the proposed backgrounds support indeed $\mathcal{N}=(2,2)$ supersymmetry. The basic idea of the argument is to show that half of the Killing spinors of $\mathrm{AdS}_3 \times \mathrm{S}^3 \times \mathbb{T}^4$ are invariant under the orbifold action (while the other half is not). The argument will be a bit sketchy, since we do not want to delve into the details of supergravity; the complete argument will be given elsewhere \cite{Eberhardt:2017uup}.

Killing spinors on $\mathrm{AdS}_3 \times \mathrm{S}^3 \times \mathbb{T}^4$ are composed of Killing spinors on $\mathrm{AdS}_3$, and Killing spinors on $\mathrm{S}^3 \times \mathbb{T}^4$. In turn, Killing spinors on $\mathrm{S}^3 \times \mathbb{T}^4$ are composed of Killing spinors on $\mathrm{S}^3$ with non-vanishing Killing constant, and parallel Killing spinors on $\mathbb{T}^4$. 
It was shown in \cite{Baer:1992aa} that Killing spinors on $\mathrm{S}^3$ are in one-to-one correspondence with Killing spinors on its Riemannian cone $\mathbb{R}^4$. Moreover, the chirality of the Killing spinor on $\mathbb{R}^4$ correlates with the sign of the Killing constant on $\mathrm{S}^3$. This sign, in turn, is mirrored by the $\mathrm{AdS}_3$ part and translates into the chirality of the corresponding supercharge in the dual CFT. 

The Killing spinors also have to obey the dilatino Killing spinor equation, which imposes a definite chirality on the $\mathrm{AdS}_3 \times \mathrm{S}^3$ part. Since Killing spinors also have a definite ten-dimensional chirality, we can impose equivalently a definite chirality on the $\mathbb{T}^4$ part. 
Thus Killing spinors on $\mathrm{AdS}_3 \times \mathrm{S}^3 \times \mathbb{T}^4$ are induced from parallel Killing spinors on 
$\mathbb{R}^4 \times \mathbb{T}^4$ with definite chirality on the $\mathbb{T}^4$, where the chirality of the Killing spinors in the dual CFT is given by the overall eight-dimensional chirality. 
The number of Killing spinors is then actually twice as large, since type IIB supergravity has two gravitinos.

Let us first discuss how the familiar cases fit into this description. $\mathbb{R}^4 \times \mathbb{T}^4$ supports $2^4=16$ parallel spinors, half of which satisfy the chirality constraint on the $\mathbb{T}^4$. Thus, the dual CFT has $\tfrac{1}{2}\times 2 \times 16=16$ supercharges. Moreover, there are equally many Killing spinors with positive eight-dimensional chirality as there are with negative chirality. Thus, we conclude that the dual CFT has eight left-moving and eight right-moving supercharges, the signature of $\mathcal{N}=(4,4)$ supersymmetry. For a more direct analysis, see for example \cite{Borsato:2014hja}.

For the case of $\mathrm{K3}$ we may perform a $\mathbb{Z}_2$ inversion orbifold on the $\mathbb{T}^4$ factor. This will again impose the same chirality constraint on the $\mathbb{T}^4$ factor, and hence will not reduce the number of Killing spinors. 
It is easy to see that the same continues to hold when we consider a $\mathbb{Z}_n$ orbifold on these coordinates as we have done above.

The orbifolds we are interested in, on the other hand, involve a $\mathbb{Z}_2$ inversion of two coordinates of the $\mathbb{R}^4$ factor, and of two  coordinates of the $\mathbb{T}^4$; this is the action of the reflection generators of the dihedral group. It imposes a chirality constraint on these directions, and hence reduces the number of Killing spinors by a factor of two. This happens independently of the eight-dimensional chirality, and thus we obtain $\mathcal{N}=(2,2)$ supersymmetry as claimed.

As an aside, we may also consider the case where we divide out the  $\mathbb{R}^4$ factor by a $\mathbb{Z}_2$ inversion. The relevant geometry is then $\mathrm{AdS}_3 \times \mathrm{SO}(3) \times \mathbb{T}^4$.  This orbifold imposes another chirality constraint on the $\mathbb{R}^4$ factor under which again only half of the Killing spinors are invariant. However, now we have fixed both the chirality on the $\mathbb{R}^4$ {\em and} on the  $\mathbb{T}^4$ factor, and hence also the full eight-dimensional chirality is  fixed. Thus all Killing spinors have the same chirality in the dual CFT, and we end up with $\mathcal{N}=(4,0)$ supersymmetry, in agreement with the analysis of \cite{Larsen:1999uk}. 

\subsection{BPS spectrum}

Next we want to determine the BPS states of supergravity. With the exception of the hyperelliptic surface of Section~\ref{subsec:hyperelliptic surface}, this will be somewhat delicate since the backgrounds are singular, and hence supergravity is not well defined. However, it makes sense in general to consider those supergravity fields on the smooth background $\mathrm{AdS}_3 \times \mathrm{S}^3 \times \mathbb{T}^4$ that are left invariant under the orbifold action. They should correspond to the untwisted sector of the $\mathrm{D}_n$ orbifold, and this is indeed what we shall find.\footnote{Note that the analysis for the hyperelliptic surface is effectively also of this kind: since the orbifold action does not have any fixed points, all BPS states arise in this manner. This is mirrored by the fact that the dual CFT also does not have any BPS states in the twisted sector.}
%

To be more precise, we shall not actually perform an honest supergravity calculation, but rather organize the KK spectrum using group theory, following the techniques of \cite{deBoer:1998kjm}. Unlike the case of $\mathrm{AdS}_3 \times \mathrm{S}^3 \times \mathrm{S}^3 \times \mathrm{S}^1$ considered in \cite{Eberhardt:2017fsi}, this actually fixes the BPS spectrum uniquely.
The calculation will be performed in two steps. First, we compactify down to six dimensions and determine the representations w.r.t.~$\mathrm{D}_n$ 
of the resulting six-dimensional fields. In a second step we then make a further KK reduction to three dimensions. We shall be considering type IIB supergravity, whose field content (in $10$ dimensions) is given in terms of $\mathfrak{so}(8)$ representations as
\be 
(\mathbf{8}_s \oplus \mathbf{8}_v)\otimes(\mathbf{8}_s \oplus \mathbf{8}_v)\ .
\ee
Reducing the theory on $\mathbb{T}^4$ simply amounts to forgetting the representations of the internal $\mathfrak{so}(4)$ of the torus. We are interested in the $\mathrm{D}_n$ representation content with respect to the $\mathrm{D}_n$ subgroup of this internal ${\rm SO}(4)$ symmetry group. Thus we have to perform the branching rules
\be 
\mathfrak{so}(8) \longrightarrow \mathfrak{so}(4) \oplus \mathfrak{so}(4) \longrightarrow \mathfrak{so}(4) \oplus \mathrm{D}_n \cong \mathfrak{su}(2) \oplus \mathfrak{su}(2) \oplus \mathrm{D}_n\ .
\ee
The relevant representations and branching rules are spelled out in Appendix~\ref{app:dihedral groups}. Under this branching we then find 
\begin{align}
\mathbf{8}_v &\longrightarrow (\mathbf{2},\mathbf{2},\mathbf{1},\mathbf{1}) \oplus (\mathbf{1},\mathbf{1},\mathbf{2},\mathbf{2}) \\
&\longrightarrow\rho_+(\mathbf{2},\mathbf{2}) \oplus 2 \rho_1 (\mathbf{1},\mathbf{1})\\
\mathbf{8}_v &\longrightarrow (\mathbf{2},\mathbf{1},\mathbf{2},\mathbf{1})\oplus(\mathbf{1},\mathbf{2},\mathbf{1},\mathbf{2}) \\
&\longrightarrow\rho_1(\mathbf{2},\mathbf{1})\oplus(\rho_+\oplus\rho_-)(\mathbf{1},\mathbf{2})\ ,
\end{align} 
where the representation labels refer to $\mathrm{D}_n$, as well as to  the six-dimensional Lorentz group $\mathfrak{su}(2) \oplus \mathfrak{su}(2)$. The six-dimensional field content then becomes 
\begin{align}
(\mathbf{8}_s \oplus \mathbf{8}_v)\otimes(\mathbf{8}_s \oplus \mathbf{8}_v)\longrightarrow &\rho_+ (\mathbf{3},\mathbf{3})\oplus 8\rho_1(\mathbf{2},\mathbf{2})\oplus(\rho_2\oplus\rho_-\oplus 2\rho_+)(\mathbf{3},\mathbf{1})\oplus 2\rho_1(\mathbf{3},\mathbf{2})\nonumber\\
&\!\!\!\!\!\!\!\!\oplus  (2\rho_+\oplus 2\rho_-)(\mathbf{2},\mathbf{3})\oplus (3\rho_+\oplus 2\rho_-)(\mathbf{1},\mathbf{3})\oplus 10\rho_1(\mathbf{1},\mathbf{2})\nonumber\\
&\!\!\!\!\!\!\!\!\oplus (4\rho_2\oplus 6\rho_-\oplus 6\rho_+)(\mathbf{2},\mathbf{1})\oplus (5\rho_2\oplus 7\rho_-\oplus 8\rho_+)(\mathbf{1},\mathbf{1})\ . \label{5.7}
\end{align} 
In a second step, we now perform the KK reduction on the three sphere $\mathrm{S}^3$ as described in \cite{deBoer:1998kjm}. However, due to the additional quotient, there are some modifications. In particular, the raising and lowering operators of the isometry group $\mathfrak{so}(4)$ transform in the $\rho_-$ representation of $\mathrm{D}_n$, which has the consequence that the different states in an $\mathfrak{so}(4)$ representation\footnote{Obviously, the orbifold group breaks the $\mathfrak{so}(4)$ symmetry. However, we may still describe the states that used to transform in an irreducible representation of $\mathfrak{so}(4)$ in this manner.}  transform in different representations of $\mathrm{D}_n$. In particular, if the highest weight state transforms in the $\rho_+$ representation, its descendants will transform either in $\rho_+$ or $\rho_-$, depending on whether an even or odd number of lowering operators have been applied. We shall continue to denote the relevant states by $(\mathbf{m},\mathbf{n})$, as in \cite{deBoer:1998kjm}. If the highest weight state transforms in some other representation $\rho$ of $\mathrm{D}_n$, we shall denote the corresponding representation by $\rho(\mathbf{m},\mathbf{n})$ (and then its states will transform in $\rho$ or $\rho\otimes \rho_-$, depending on how many lowering operators have been applied).

The analysis is then fairly straightforward, except that it is sometimes not easy to see whether the highest weight state of a representation is even or odd w.r.t.~$\mathrm{D}_n$. This question is easily answered for bosons, since the wavefunction of a boson with even spin is even under a rotation by $\pi$, while it is odd for odd spin. However, for the fermions this question is more delicate. As it turns out, the ambiguity does not actually affect the final answer since all the representations that appear for the fermions satisfy $\rho \otimes \rho_- \cong \rho$. Upon adding the various contributions, one then obtains
\begin{align}
&\bigoplus_\mathbf{m}\rho_+ (\mathbf{m},\mathbf{m\pm 4})\oplus (2\rho_1\oplus 2\rho_+\oplus 2\rho_-)(\mathbf{m},\mathbf{m\pm 3}) \nonumber \\
& \qquad \oplus (3\rho_+\oplus 7\rho_-\oplus 8\rho_1\oplus \rho_2)(\mathbf{m},\mathbf{m\pm 2}) \oplus (10\rho_+\oplus 10\rho_-\oplus 14\rho_1\oplus 4\rho_2)(\mathbf{m},\mathbf{m\pm 1}) \nonumber\\[4pt]
&\qquad\oplus (16\rho_+\oplus 10\rho_-\oplus 16\rho_1\oplus 6\rho_2)(\mathbf{m},\mathbf{m})\ . \label{CFT_sugra_spectrum}
\end{align} 
This is the complete supergravity spectrum. We can determine from this the BPS spectrum by fitting the states into modified $\mathcal{N}=4$ multiplets, which we shall denote by $(\mathbf{m},\mathbf{n})_\mathrm{S}$; their structure is described in more detail in Appendix~\ref{app:N4N2}. It is a strong consistency check that this is possible, and we find 
\be 
\bigoplus_\mathbf{m} \rho_- (\mathbf{m},\mathbf{m\pm 2})_\mathrm{S} \oplus 2\rho_1(\mathbf{m},\mathbf{m\pm 1})_\mathrm{S} \oplus (3\rho_+ \oplus \rho_- \oplus \rho_2)(\mathbf{m},\mathbf{m})_\mathrm{S}\ . \label{sugra_BPS_spectrum}
\ee
This reproduces \eqref{symmetric orbifold untwisted BPS spectrum}. (There are low lying exceptions in both formulae, but they also match precisely.)

\section{String theory} \label{sec:string theory}

In this section we shall analyze the same backgrounds using a stringy world-sheet description. We shall consider the pure NS-NS flux case for which the AdS$_3 \times {\rm S}^3$ factor can be described (before orbifolding) by an $\mathfrak{sl}(2,\mathbb{R}) \oplus \mathfrak{su}(2)$  WZW model. The remaining $\mathbb{T}^4$ (before orbifolding) is simply described by free fields, and the orbifold acts on the world-sheet fields in a geometrical manner. Using this approach we shall determine the spacetime BPS spectrum and the supergravity elliptic genus of \cite{deBoer:1998us}, and compare them to the dual CFT predictions. We shall mainly be interested in the situation where the levels of the WZW models are large; in particular, we shall only consider the unflowed sector of the 
$\mathfrak{sl}(2,\mathbb{R}) $ WZW model. (More details about the construction of the relevant WZW model can be found in \cite{Ferreira:2017pgt} and references therein; we shall follow the conventions of \cite{Eberhardt:2017pty}.)

%
%
%

In order to fix notation, let us denote the $\mathfrak{sl}(2,\mathbb{R})$ fermions by $\psi^\pm$, the $\mathfrak{su}(2)$ fermions by $\chi^\pm$, and the four torus fermions by $\lambda^i$, $i=1,\dots,4$. Note that we have eliminated two fermions due to the physical state conditions. These fermions sit in representations of $\mathrm{D}_n$ as
\begin{align}
&\psi^\pm\ : \quad 2\rho_+ \\
&\chi^\pm\ : \quad 2\rho_- \\
&\lambda^i\ \ : \quad 2 \rho_1\ .
\end{align} 
The same is, of course, true for the bosons, but we shall only need their zero modes. Since the raising and lowering operators of the $\mathfrak{su}(2)$ algebra (the bosonic analogues of $\chi^\pm$) transform in the representation $\rho_-$, we obtain exactly the same structure of $\mathfrak{su}(2)$ representations we have discussed above.

\subsection{Untwisted sector}

The analysis in the untwisted NS sector is straightforward. Suppose the ground states transform in the representation $\mathbf{m}$ of $\mathfrak{su}(2)$ --- to make contact with the supergravity and CFT answer, we shall use the same notation as de Boer, see \cite{deBoer:1998kjm}, \ie ${\bf m}$ is the $m$-dimensional representation of $\mathfrak{su}(2)$ with spin $j=(m-1)/2$.\footnote{Recall that the different vectors of ${\bf m}$ transform in either $\rho_+$ or $\rho_-$; the highest weight state transforms in $\rho_+$, and each time the weight is reduced by one (via the action of $J^-$) the representation flips from $\rho_+$ to $\rho_-$ or vice versa, see also the comment below eq.~(\ref{5.7}). Thus there are $\lfloor \frac{m}{2}\rfloor$ states in $\rho_-$, while the remaining states transform in $\rho_+$.} Then the massless states that appear at excitation level $1/2$ transform as 
\be 
(2\rho_++2\rho_1)\mathbf{m}\oplus\rho_-(\mathbf{m+2})\oplus\rho_-(\mathbf{m-2})\ . \label{NS_sector_untwisted_field_content}
\ee
They contain the BPS states
\be 
\rho_+\mathbf{m} \oplus \rho_-(\mathbf{m+2})\ ,\label{NS_sector_untwisted_BPS_content}
\ee
since we can either apply one $\mathfrak{sl}(2,\mathbb{R})$ fermion or one $\mathfrak{su}(2)$ fermion to obtain a BPS state.  

The R sector analysis is a bit more subtle. The Ramond ground states (before GSO projection) transform in the $4\cdot (\mathbf{2},\mathbf{2})$ of $\mathfrak{sl}(2,\mathbb{R}) \oplus \mathfrak{su}(2)$. Here, the $(\mathbf{2},\mathbf{2})$ is the contribution from the first six coordinates, while the $4$ accounts for the multiplicity coming from the torus coordinates. 
The torus fermionic zero modes transform in the spinor representation 
$(\mathbf{2},\mathbf{1}) \oplus (\mathbf{1},\mathbf{2})$  of the internal $\mathfrak{su}(2) \oplus \mathfrak{su}(2)\cong\mathfrak{so}(4)$ of the torus. Upon
branching to $\mathrm{D}_n$, this becomes $\rho_1 \oplus \rho_- \oplus \rho_+$.
Thus the Ramond ground states transform in the representation $(\rho_1 \oplus \rho_- \oplus \rho_+)(\mathbf{2},\mathbf{2})$. 

The R sector contains a BPS multiplet (at excitation level zero), and the BPS representation content is $\rho_1 \oplus \rho_- \oplus \rho_+$ before applying the GSO projection. After the GSO projection, we should either retain $\rho_1$ or $\rho_+\oplus\rho_-$. The case of $\mathrm{K3}$\footnote{For $\mathrm{K3}=\mathbb{T}^4/\mathbb{Z}_2$, the orbifold group is $\mathbb{Z}_2 \subset \mathrm{D}_2$, and $\rho_1$ branches to twice the non-trivial representation of $\mathbb{Z}_2$, whereas both $\rho_\pm$ branch to the trivial representation.} tells us the correct result: the choice giving the right $\mathrm{K3}$ BPS spectrum is $\rho_1$. Thus, we conclude that in general the BPS content of the R sector is
\be 
\rho_1 (\mathbf{m+1})\ .\label{R_sector_untwisted_BPS_content}
\ee
Together with the contributions from the untwisted NS sector and combining left- and right-movers we then obtain for the entire BPS spectrum 
\begin{align}
\bigoplus_\mathbf{m} &(\rho_+\mathbf{m} \oplus \rho_-(\mathbf{m+2}) \oplus \rho_1 (\mathbf{m+1}))^2 \nonumber\\
&\cong \bigoplus_\mathbf{m} \rho_-(\mathbf{m},\mathbf{m\pm 2}) \oplus 2\rho_1(\mathbf{m},\mathbf{m\pm 1}) \oplus (\rho_2 \oplus\rho_-\oplus 3\rho_+)(\mathbf{m},\mathbf{m}) \ ,
\end{align} 
which is precisely \eqref{sugra_BPS_spectrum}. It is also a simple matter to confirm the low lying exceptions from the point of view of string theory. Thus we have reproduced the supergravity spectrum from the untwisted sector of the world-sheet description. 


 \subsection{Twisted sectors}
 
\subsection*{The twisted sector for $\mathrm{AdS}_3 \times (\mathrm{S}^3 \times \mathbb{T}^2)/\mathbb{Z}_2 \times \mathbb{T}^2$}

The analysis in the twisted sector is more complicated, and we shall first concentrate on the case of $\mathrm{AdS}_3 \times (\mathrm{S}^3 \times \mathbb{T}^2)/\mathbb{Z}_2 \times \mathbb{T}^2$.
Since the orbifold generator also acts non-trivially on the ${\rm S}^3$ factor (namely by rotation by $180$ degrees), the twist will also affect the affine $\mathfrak{su}(2)$ algebra. In fact, in terms of characters, the insertion of the twist operator implies that the $\mathfrak{su}(2)$ part of the partition function is 
\be 
\text{\scalebox{.7}{1}}\underset{-1}{\raisebox{-5pt}{\text{\scalebox{2}{$\square$}}}}=Z(z+\tfrac{1}{2}, \tau)\ .
\ee
The corresponding twisted sector then has the character 
\be 
\text{\scalebox{.7}{$-1$}}\underset{1}{\raisebox{-5pt}{\text{\scalebox{2}{$\square$}}}}=Z\left(\frac{z+\frac{1}{2}\tau}{\tau}, \tau\right)=\exp\left(\tfrac{\pi i}{2\tau}(k-2)(z+\tfrac{\tau}{2})^2\right)Z(z+\tfrac{1}{2}\tau, \tau)\ ,
\ee
where we have used the modular properties of the $\mathfrak{su}(2)$ partition function in the last step, see \eg \cite{Gaberdiel:2012yb}. Formally, this has the same form as the character that is obtained by spectral flow by half a unit. In fact, this was to be expected since one unit of spectral flow corresponds to a rotation by $2\pi$, whereas we are here only rotating by $\pi$. Thus we can describe the $\mathfrak{su}(2)$ part of the twisted sector by simply spectrally flowing by half a unit. 

The other building blocks of the twisted sector are easier to describe: 
orbifolding a boson, \ie $\mathrm{S}^1/\mathbb{Z}_2$, gives a ground state energy of $\tfrac{1}{16}$ in the twisted sector (and a half-integer moded boson). Similarly, orbifolding a NS fermion gives $\tfrac{1}{16}$ (and an integer moded fermion), while a R fermion gives a ground state energy of $-\tfrac{1}{16}$ (and becomes half-integer moded). 
 
\subsubsection*{BPS states in NS sector}

With these preparations at hand, we can now look for spacetime BPS states in the $\mathbb{Z}_2$ twisted sector --- we shall be somewhat brief in the following as the calculation proceeds very similarly to Section~4 of  \cite{Eberhardt:2017pty}.
Let us denote the ground state spins  of $\mathfrak{sl}(2,\mathbb{R})_{k+2}$ and $\mathfrak{su}(2)_{k-2}$ by $j_0$ and $\ell_0$, respectively (where $\ell_0$ is evaluated before the half-unit spectral flow). As in the untwisted sector we can apply  one $\mathfrak{sl}(2,\mathbb{R})$ fermion (still half-integer moded) on the ground state to lower the $\mathfrak{sl}(2,\mathbb{R})$ spin by one unit. This is however the only possibility to obtain a BPS state, since we cannot use the $\mathfrak{su}(2)$ fermion in the same way, since it is twisted. (On the spectrally flowed ground state, the twisted $\mathfrak{su}(2)$ fermions and the two torus fermions generate a $2^2=4$-dimensional spinor representation $2 \cdot \mathbf{2}$ of $\mathfrak{su}(2)_k$. The GSO projection allows only an odd number of fermions in total, so we obtain only the representation $\mathbf{2}$.) We consider the highest weight state of this $\mathfrak{su}(2)$ representation, since it has the potential to be a BPS state. Its true spacetime spins are 
\be 
j=j_0-1\ , \quad \ell=\ell_0+\frac{k-2}{4}+\frac{1}{2}=\ell_0+\frac{k}{4}\ , \label{true_spins_NS1}
\ee
where the expression for $\ell$ can be obtained by flowing by half a unit in the supersymmetric $\mathfrak{su}(2)_k$ current. The mass shell condition for this state is on the other hand\footnote{The first two terms come from the Casimirs of $\mathfrak{sl}(2,\mathbb{R})$ and $\mathfrak{su}(2)$, and the next two from the half-unit spectral flow of $\mathfrak{su}(2)_{k-2}$. The $\tfrac{1}{8}$ is the ground state energy of the two twisted bosons of the $\mathbb{T}^2$. Finally the $\tfrac{1}{4}$ comes from the ground state energy of the four twisted fermions and the $\tfrac{1}{2}$ because the excitation level is $\tfrac{1}{2}$.}
\be \label{BPS1}
-\frac{j_0(j_0-1)}{k}+\frac{\ell_0(\ell_0+1)}{k}+\frac{1}{2}\ell_0+\frac{k-2}{16}+\frac{1}{8}+\frac{1}{4}+\frac{1}{2}=\frac{1}{2}\ ,
\ee
which upon insertion of \eqref{true_spins_NS1} tells us that $j=\ell$, and hence that the state is indeed BPS.
Because of \eqref{true_spins_NS1} this solution only exists for $\ell \ge \frac{k}{4}$. Below this value, we have to use the $-\tfrac{1}{2}$ spectral flow of $\mathfrak{su}(2)_{k-2}$ to get a simple result. Here, no $\mathfrak{sl}(2)$ fermion has to be applied for the BPS state, and the true spacetime spins are given by
\be \label{BPS2}
j=j_0\ , \quad \ell=\ell_0-\frac{k-2}{4}+\frac{1}{2}=\ell_0-\frac{k}{4}+1\ .
\ee
This can also be reinterpreted as a supersymmetric spectral flow of the state where one NS moded $\mathfrak{su}(2)$ fermion was applied to the ground state. This solution works up to $\tfrac{k-2}{2}-\tfrac{k}{4}+1=\tfrac{k}{4}$, complementing the other series (that starts at $\tfrac{k}{4}$). Consequently, the state at spin $\tfrac{k}{4}$ occurs twice.



\subsubsection*{BPS states in R sector}

The R sector analysis works similarly. We again distinguish the cases where we use the spectral flow with $w=\tfrac{1}{2}$ or $w=-\tfrac{1}{2}$ in the $\mathfrak{su}(2)$ sector. Let us start with $w=\tfrac{1}{2}$. Then we apply after the spectral flow one NS moded  $\mathfrak{sl}(2,\mathbb{R})$ fermion on the state. (These fermions transform in the representation $2 \cdot \mathbf{2}$ of $\mathfrak{sl}(2,\mathbb{R})$, which is cut down to $\mathbf{2}$ by the GSO projection.) This time, only the lowest weight state of this representation has a chance to be BPS. For this state, the true spins are given by
\be 
j=j_0-\frac{1}{2}\ , \quad \ell=\ell_0+\frac{k-2}{4}+1=\ell_0+\frac{1}{2}+\frac{k}{4}\ , \label{true_spins_R1}
\ee
which can again be interpreted as a spectral flow of the BPS state in the untwisted sector. The corresponding mass shell condition is
\be 
-\frac{j_0(j_0-1)}{k}+\frac{\ell_0(\ell_0+1)}{k}+\frac{1}{2}\ell_0+\frac{k-2}{16}-\frac{1}{8}+\frac{1}{2}=0\ ,
\ee
which yields indeed $j=\ell$, so the state is again BPS. Again, the state exists only for $j=\ell \ge \tfrac{k+2}{4}$.

Finally, if we use the spectral flow $w=-\tfrac{1}{2}$, we should not apply any further fermions, so the true spins are
\be 
j=j_0-\frac{1}{2}\ , \quad \ell=\ell_0-\frac{k-2}{4}=\ell_0+\frac{1}{2}-\frac{k}{4}\ , \label{true_spins_R2}
\ee
which looks again like a supersymmetric spectral flow of the untwisted BPS state. The mass shell condition
\be 
-\frac{j_0(j_0-1)}{k}+\frac{\ell_0(\ell_0+1)}{k}-\frac{1}{2}\ell_0+\frac{k-2}{16}-\frac{1}{8}=0\ ,
\ee
gives again $j=\ell$. This solution is valid up to $\ell=\tfrac{k-2}{2}+\frac{1}{2}-\frac{k}{4}=\tfrac{k-2}{4}$, so the BPS state $\tfrac{k}{4}$ is missing. 
This is actually required, given that it appears twice in the NS sector. We have then precisely two BPS states at every spin (with the exception of the lowest spin, which comes from the R sector). It is a nice consistency check that the spectrum fits into $\mathcal{N}=2$ multiplets; below $j=\tfrac{k}{4}$, the highest weight state comes from the R sector, above from the NS sector. 
\subsubsection*{The full twisted sector}

We have seen above that the NS-NS, the NS-R, the R-NS and the R-R sector each contribute one BPS state; these different states arrange themselves into the diamond
\be \label{diamond1}
\begin{tabular}{ccc}
& 1 & \\
1 & & 1 \\
& 1 &
\end{tabular}\ .
\ee
Since moreover, the orbifold action has four fixed points on the two-torus, they are four-fold degenerate. In order to reproduce the BPS spectrum of the symmetric orbifold associated to $\mathrm{D}_1^{(1)}$, see in particular Fig.~\ref{tab:torus quotients BPS spectra}, we need that these twisted sector contributions obey $j=\ell \in \frac{1}{2}\mathbb{Z} + \frac{1}{4}$ --- recall that the twisted sector Hodge numbers appear for half-integer $(p,q)$.  Given the form of eqs.~(\ref{BPS1}), (\ref{BPS2}), (\ref{true_spins_R1}) and (\ref{true_spins_R2}), it follows that this is only the case provided that $k$ is {\em odd}.

The condition that $k$ has to be odd also follows from spacetime supersymmetry. Recall that spacetime supersymmetry requires that the world-sheet theory is ${\cal N}=2$ and has an integral $\mathfrak{u}(1)$ spectrum \cite{Banks:1987cy}. By construction the world-sheet theory is ${\cal N}=2$ before orbifolding since each factor of the world-sheet description 
\be 
\mathrm{AdS}_3 \times \mathrm{S}^3 \times \mathbb{T}^4 \cong \frac{\mathfrak{sl}(2,\mathbb{R})^{(1)}_k}{\mathfrak{u}(1)^{(1)}}\oplus \frac{\mathfrak{su}(2,\mathbb{R})^{(1)}_k}{\mathfrak{u}(1)^{(1)}} \oplus (\mathfrak{u}(1)^{(1)})^{\oplus 6}
\ee
is. Here, the superscript (1) refers to the fact that we are considering the $\mathcal{N}=1$ supersymmetric affine algebra. Thus, in addition to the bosonic WZW models, we have an additional set of real fermions transforming in the adjoint representation. For more details on this construction, see {\it e.g.}~\cite{Gaberdiel:2013vva}. The $\mathfrak{u}(1)$ charge of the different factors are also integer valued.\footnote{There is a factor of two involved in the conventions when going from the $\mathfrak{su}(2)$ spins to the $\mathfrak{u}(1)$ charge.} When taking the orbifold, the same is clearly true for the untwisted sector. However, we saw above that the $\mathfrak{u}(1)$ charge of the $\mathrm{S}^3$ part becomes half-integer valued if $k$ is odd and integer-valued if $k$ is even. Furthermore, the $\mathfrak{u}(1)$ charge of the $\mathbb{T}^2$ factor is always half-integer in the twisted sector. Thus, the complete $\mathfrak{u}(1)$ charge will only be integer-valued if $k$ is odd. 
We therefore conclude that, provided $k$ is odd, string theory on $\mathrm{AdS}_3 \times \bigl( \mathrm{S}^3 \times \mathbb{T}^2 \bigr)/\mathbb{Z}_2 \times \mathbb{T}^2$ is supersymmetric, and its BPS spectrum matches  precisely with that of the symmetric orbifold based on $\mathrm{D}_1^{(1)}$.

\subsection*{The twisted sector for $\mathrm{AdS}_3\times (\mathrm{S}^3 \times \mathbb{T}^2\times \mathbb{T}^2)/(\mathbb{Z}_2 \times \mathbb{Z}_2)$}

We can also compute fairly directly the twisted sector contribution for the case of $\mathrm{AdS}_3\times (\mathrm{S}^3 \times \mathbb{T}^2\times \mathbb{T}^2)/(\mathbb{Z}_2 \times \mathbb{Z}_2)$. In this case, there are three twisted sectors: we can either twist with only one of the $\mathbb{Z}_2$'s or with the diagonal $\mathbb{Z}_2$. (In the language of Section~\ref{sec:elliptic genus}, they correspond to the contributions UT and TT, respectively.) The twist of the diagonal $\mathbb{Z}_2$ is precisely the same as for $\mathrm{K3}$, and thus we obtain $16$ additional scalar BPS states for $(p,q)=(n,n)$ with $n=1,2,\dots$. 

More interesting is the twist of only one $\mathbb{Z}_2$, \ie the UT sector. The BPS states of that twisted sector are again described by (\ref{diamond1}), and they are again $4$-fold degenerate, but invariance under the full orbifold group $\mathbb{Z}_2 \times \mathbb{Z}_2$ removes the two middle states. There are two twisted sectors of this kind (UT and UT), and thus we get $8$ such states for each half-integer Hodge number, \ie for $(p,q)=(n+\frac{1}{2},n+\frac{1}{2})$ with $n=0,1,2,\ldots$. (We are assuming here again that $k$ is odd so that the spins coming from the TU and UT sectors are quarter-integer.) 
The complete BPS spectrum of $\mathrm{AdS}_3\times (\mathrm{S}^3 \times \mathbb{T}^2\times \mathbb{T}^2)/(\mathbb{Z}_2 \times \mathbb{Z}_2)$ for $k$ odd is then given by overlapping diamonds of the form
\be 
\begin{tabular}{ccc}
& 1 & \\
0& 8 &0 \\
0& 18 &0 \\
0& 8 &0 \\
& 1 &
\end{tabular}\ .
\ee 
This is then in perfect agreement with the BPS spectrum given in Section~\ref{sec:CFT}.

\subsection*{The twisted sector for $\mathrm{AdS}_3\times (\mathrm{S}^3 \times \mathbb{T}^4)/\mathrm{D}_n$}

Actually, the previous analysis generalizes fairly directly also to the other $\mathrm{D}_n$ orbifolds. The twisted sector states are either associated to a cyclic rotation generator, or to a reflection generator. 
The analysis for the reflection generators\footnote{For odd $n$, all reflection generators sit in the same conjugacy class of  ${\rm D}_n$, while for even $n$ there are two conjugacy classes. Both, however, give the same contribution.} works as described above for the case of $\mathrm{AdS}_3 \times (\mathrm{S}^3 \times \mathbb{T}^2)/\mathbb{Z}_2 \times \mathbb{T}^2$. On the other hand, the cyclic rotation generators all appear in standard K3 orbifolds, and one can use K3 results for them (as we have also done just now for the case of $\mathrm{AdS}_3\times (\mathrm{S}^3 \times \mathbb{T}^2\times \mathbb{T}^2)/(\mathbb{Z}_2 \times \mathbb{Z}_2)$).
As a consequence the string theory calculation essentially just mirrors the CFT calculation --- one only needs to  keep track of which states survive the orbifold projection, but this works equally on both sides. Thus the string theory spectrum is comprised of overlapping Hodge diamonds, and the relevant Hodge diamonds are precisely those of the seed theory discussed in Section~\ref{sec:CFT}, see Tables~\ref{tab:torus quotients BPS spectra} and \ref{tab:torus quotients BPS spectra discrete torsion}.

\subsection{Elliptic genus}

 \subsection*{The elliptic genus for $\mathrm{AdS}_3\times (\mathrm{S}^3 \times \mathbb{T}^2\times \mathbb{T}^2)/(\mathbb{Z}_2 \times \mathbb{Z}_2)$}
 
 Finally we want to match the CFT elliptic genus from supergravity/string theory. We shall determine first the elliptic genus for the case of $\mathrm{AdS}_3\times (\mathrm{S}^3 \times \mathbb{T}^2\times \mathbb{T}^2)/(\mathbb{Z}_2 \times \mathbb{Z}_2)$, and then explain how the calculation generalizes to the other cases. (For the case of $\mathrm{AdS}_3\times (\mathrm{S}^3 \times \mathbb{T}^2 )/\mathbb{Z}_2\times \mathbb{T}^2$ the elliptic genus vanishes because of the $\mathbb{T}^2$ factor.)  
 
 As in \cite{deBoer:1998us} we shall consider the supergravity limit $k \to \infty$, in which most states acquire infinite mass and disappear from the spectrum. In the untwisted sector, the states that survive just make up the supergravity spectrum, but string theory provides additionally also a description of the twisted sectors.  The states that contribute to the supergravity elliptic genus are in general quarter BPS states; 
for $\mathrm{AdS}_3 \times\mathrm{S}^3 \times \mathbb{T}^4$ they are simply given by a BPS state for the left-movers and a descendant in the global supergravity multiplet for the right-movers \cite{deBoer:1998us}. For the present case, the situation is essentially the same, except that we have to keep track of the representation content with respect to the orbifold group (since, in the end, only the invariant states survive).  

Let us start with analyzing the untwisted sector. There, only every second BPS state survives the orbifold projection, and thus the relevant character is
 \be 
 Z^{\mathrm{sp}}_\mathrm{UU}(z,\tau)=\sum_{\ell=0,\, \ell \in \frac{1}{2}\mathbb{Z}}^\infty \left(\chi^{\mathcal{N}=4}_\ell(z,\tau)+2\chi^{\mathcal{N}=4}_{\ell+\frac{1}{2}}(z,\tau)+\chi^{\mathcal{N}=4}_{\ell+1}(z,\tau)\right)\ ,
 \ee
 where, `sp' refers to fact that we are considering the single particle character. 
 For the TT contribution, the effect of the orbifold is that we have to replace the usual ${\cal N}=4$ characters with their modified versions, see eq.~(\ref{modchar})
 \be 
 Z^{\mathrm{sp}}_\mathrm{TT}(z,\tau)=16\sum_{\ell=0,\, \ell \in \frac{1}{2}\mathbb{Z}}^\infty\widetilde{\chi}^{\mathcal{N}=4}_{\ell+\frac{1}{2}}(z,\tau)\ .
 \ee
 Finally for the UT contribution, we only have an $\mathcal{N}=2$ multiplet, not the remnants of an $\mathcal{N}=4$ multiplet. Thus, we have
 \be 
 Z^{\mathrm{sp}}_\mathrm{UT}(z,\tau)=\sum_{\ell=0,\, \ell \in \frac{1}{2}\mathbb{Z}}^\infty \left(8\chi^{\mathcal{N}=2}_{\ell+\frac{1}{4}}(z,\tau)+8\chi^{\mathcal{N}=2}_{\ell+\frac{3}{4}}(z,\tau)\right)\ .
 \ee
One may then easily perform the series expansion of the characters, and after some non-trivial cancellations, one finds\footnote{In order to match with the conventions of the dual CFT calculation, we have to multiply the $\mathfrak{su}(2)$ quantum numbers by a factor of two.}
 \begin{align}
 Z^{\mathrm{sp}}_\mathrm{UU}(2z,\tau)&=-5-\frac{2+q^{1/2}(y^2+y^{-2})}{1-q}+\frac{4}{1-q^{1/2}y^2}+\frac{4}{1-q^{1/2}y^{-2}}\ , \\
 Z^{\mathrm{sp}}_\mathrm{UT}(2z,\tau)&=-\frac{8(q^{1/4}y+q^{3/4}y^{-1})}{1-q}+\frac{16q^{1/4}y}{1-q^{1/2}y^2}\ , \\
 Z^{\mathrm{sp}}_\mathrm{TT}(2z,\tau)&=-16-\frac{16}{1-q}+\frac{16}{1-q^{1/2}y^2}+\frac{16}{1-qy^{-4}}\ .
 \end{align}
Let us denote by $g_*(m,\ell)$ the coefficients of these functions, where $*$ stands either for UU, UT or TT; explicit formulae for these can be obtained from the  above expressions. One finds that the result agrees precisely with Fig.~\ref{tab:sum-of-coefficients}, \ie 
\be
g_*(m,l)=\sum_{n>0} d_*(n,m,l)\ .
\ee
This provides a fairly non-trivial confirmation of our duality proposal, at least for this case. 

\subsection*{The elliptic genus for $\mathrm{AdS}_3\times (\mathrm{S}^3 \times \mathbb{T}^4)/\mathrm{D}_n$}

We can generalize the above analysis also to the other orbifolds. As we have shown above, the 
BPS spectrum is given by overlapping Hodge diamonds. The integer Hodge diamond entries stem always from the untwisted sector (UU) or the twisted sectors with respect to some elements of the $\mathbb{Z}_n \subset \mathrm{D}_n$ orbifold group (TT). Thus, they always come in modified $\mathcal{N}=4$ representations. Half-integer Hodge numbers, on the other hand, arise in the twisted sectors w.r.t.~reflections of the dihedral group (UT), and they only organize themselves in  $\mathcal{N}=2$ representations. For the UU contribution to the single particle elliptic genus, we find
\begin{align}
Z^\mathrm{sp}_\mathrm{UU}(z,\tau)&=\sum_{\ell=0, \, \ell \in \frac{1}{2}\mathbb{Z}}^\infty \Big((\rho_++\rho_--\rho_1)\widetilde{\chi}_\ell^{\mathcal{N}=4}(z,\tau)+(\rho_++\rho_--2\rho_1+\rho_2) \widetilde{\chi}_{\ell+\frac{1}{2}}^{\mathcal{N}=4}(z,\tau)\nonumber\\
&\qquad\qquad +(\rho_++\rho_--\rho_1)\widetilde{\chi}_{\ell+1}^{\mathcal{N}=4}(z,\tau)\Big)\\
&=\sum_{\ell=0,\,  \ell \in \frac{1}{2}\mathbb{Z}}^\infty \Big((\rho_+-\rho_1)(\chi_\ell^{\mathcal{N}=4}(z,\tau)+\chi_{\ell+1}^{\mathcal{N}=4}(z,\tau))\nonumber\\
&\qquad\qquad+(\rho_+-2\rho_1+\rho_2)\chi^{\mathcal{N}=4}_{\ell+\frac{1}{2}}(z,\tau)\Big)\ .
\end{align}
Here, we have used that all representations that appear satisfy $\rho \otimes \rho_-  \cong \rho$, and hence that the expression reduces to standard $\mathcal{N}=4$ characters. We can furthermore express this in terms of the untwisted contribution to the Hodge diamond as
\begin{align} 
Z^\mathrm{sp}_\mathrm{UU}(z,\tau)&=\sum_{\ell=0,\ \ell \in \frac{1}{2}\mathbb{Z}}^\infty \Big((1-h_{0,1})(\chi_\ell^{\mathcal{N}=4}(z,\tau)+\chi_{\ell+1}^{\mathcal{N}=4}(z,\tau))\nonumber\\
&\qquad\qquad\qquad +(h_{1,1}^\mathrm{U}-2h_{1,0})\chi^{\mathcal{N}=4}_{\ell+\frac{1}{2}}(z,\tau)\Big)\ .
\end{align}
Next we determine the contribution of the TT sector. Note that this sector only exists when we are orbifolding with $\mathrm{D}_n$ for $n \ge 2$; we will therefore consider $n \ge 2$ from now on. Given that the TT sector also appears in the corresponding K3 orbifold, we know that the contribution from the cyclically twisted sectors ensures that the middle Hodge number of K3 adds up to 20. (This number is $16$ for $n=2$ and $18$ for $n \ge 3$.) 
However, not all of these $20-2h_{1,1}^\mathrm{U}$ chiral states are invariant under the reflection $\mathbb{Z}_2$. To end up with the correct Hodge diamond of \eqref{quotient_untwisted_Hodge_diamond}, there have to be $h_{1,1}-h_{1,1}^\mathrm{U}$ states transforming in $\rho_+$, and $20-h_{1,1}-h_{1,1}^\mathrm{U}$ states transforming in $\rho_-$. Thus, we conclude that the contribution of the TT sector to the elliptic genus is given by
\begin{align} 
Z^\mathrm{sp}_\mathrm{TT}(z,\tau)&=\sum_{\ell=0,\, \ell \in \frac{1}{2}\mathbb{Z}}^\infty \Big((h_{1,1}-h_{1,1}^\mathrm{U})\widetilde{\chi}^{\mathcal{N}=4}_{\ell+\frac{1}{2}}(z,\tau)
\nonumber\\
&\qquad\qquad
+(20-h_{1,1}-h_{1,1}^\mathrm{U})(\chi^{\mathcal{N}=4}_{\ell+\frac{1}{2}}(z,\tau)-\widetilde{\chi}^{\mathcal{N}=4}_{\ell+\frac{1}{2}}(z,\tau))\Big)\\ 
&=\sum_{\ell=0,\, \ell \in \frac{1}{2}\mathbb{Z}}^\infty\Big((20-h_{1,1}-h_{1,1}^\mathrm{U})\chi^{\mathcal{N}=4}_{\ell+\frac{1}{2}}(z,\tau)+(2h_{1,1}-20)\widetilde{\chi}^{\mathcal{N}=4}_{\ell+\frac{1}{2}}(z,\tau)\Big)\ . \nonumber
\end{align}
The contribution from the UT sector is much easier to determine, since the chiral states are only associated to $\mathcal{N}=2$ multiplets, and we obtain
 \be 
 Z^{\mathrm{sp}}_\mathrm{UT}(z,\tau)=\sum_{\ell=0,\, \ell \in \frac{1}{2}\mathbb{Z}}^\infty \left(h_{\frac{1}{2},\frac{1}{2}}-h_{\frac{1}{2},\frac{3}{2}}\right)\left(\chi^{\mathcal{N}=2}_{\ell+\frac{1}{4}}(z,\tau)+\chi^{\mathcal{N}=2}_{\ell+\frac{3}{4}}(z,\tau)\right)\ .
 \ee
Note that for $n=1$ the supergravity elliptic genus vanishes identically, in agreement with the proposed CFT. So let us again assume $n \ge 2$. We note that upon adding up the contributions of the different sectors, the result does no longer depend explicitly on $h_{1,1}^\mathrm{U}$. In fact, using also that $h_{0,1}=0$, it depends only on $h_{1,1}$ and the combination $h_{\frac{1}{2},\frac{1}{2}}-h_{\frac{1}{2},\frac{3}{2}}$. We can express these quantities in terms of the parameter $a$ which we introduced in Section~\ref{sec:ell},
\be 
h_{1,1}=14+4a\ , \quad h_{\frac{1}{2},\frac{1}{2}}-h_{\frac{1}{2},\frac{3}{2}}=4a+4\ .
\ee
Thus, we can give the full supergravity elliptic genus in terms of the parameter $a$ as
\begin{align}
Z^{\mathrm{sp}}(z,\tau)&=\sum_{\ell=0,\ \ell \in \frac{1}{2}\mathbb{Z}}^\infty \Big(\chi_\ell^{\mathcal{N}=4}(z,\tau)+\chi_{\ell+1}^{\mathcal{N}=4}(z,\tau)+(6-4a)\chi^{\mathcal{N}=4}_{\ell+\frac{1}{2}}(z,\tau)\nonumber\\
&\qquad+(8a+8)\widetilde{\chi}^{\mathcal{N}=4}_{\ell+\frac{1}{2}}(z,\tau)+(4a+4)(\chi^{\mathcal{N}=2}_{\ell+\frac{1}{4}}(z,\tau)+\chi^{\mathcal{N}=2}_{\ell+\frac{3}{4}}(z,\tau))\Big)\ .
\end{align}
It is then a simple matter to determine the expansion of this expression to extract the coefficients $g(m,\ell)$. The result agrees precisely with Table~\ref{tab:coefficients general elliptic genus}.
\section{Conclusions} \label{sec:conclusions}

In this paper we have conjectured new ${\rm AdS}_3/{\rm CFT}_2$ dualities with $\mathcal{N}=(2,2)$ supersymmetry. The relevant string backgrounds are ${\rm D}_n$ orbifolds  ${\rm AdS}_3 \times ({\rm S}^3\times \mathbb{T}^4)/{\rm D}_n$, where $\mathrm{D}_n$ acts on the torus, and the inversion generators of ${\rm D}_n$  also rotate the ${\rm S}^3$ by 180 degrees. The CFT duals to these theories lie on the same moduli space as the  symmetric orbifold of $\mathbb{T}^4/{\rm D}_n$. We have checked that the BPS spectrum and the  elliptic genus matches between the two descriptions.

It would be interesting to consider similarly quotients of the background with large $\mathcal{N}=4$ supersymmetry, ${\rm AdS_3}\times {\rm S}^3 \times {\rm S}^3 \times {\rm S}^1$, and investigate whether an analogous analysis applies in that case. One may also study quotients of the background ${\rm AdS}_3 \times ({\rm S}^3\times \mathbb{T}^4)/{\rm G}$ with the goal of constructing holographic dualities with $\mathcal{N}=(2,0)$ supersymmetry; similar constructions have already been considered in 
\cite{Larsen:1999uk, Hohenegger:2008du, Gava:2001ne}, but the advantage of the present setting is that we have retained $\mathcal{N}=(2,2)$ supersymmetry, rather than just ${\cal N}=(2,0)$. In particular, the dual CFTs are thus much more constrained.

It would also be important to understand how the string backgrounds can be obtained as limits of brane configurations. In particular, for the background with pure NS-NS flux, the world-sheet analysis (based on a WZW model) implied that the level of the $\mathfrak{sl}(2,\mathbb{R})$ algebra has to be odd in order to preserve spacetime supersymmetry. This should translate into the statement that a certain brane charge needs to be odd, and it would be interesting to understand this effect also from other perspectives. 

Finally, it would be worthwhile to investigate whether one can identify the string duals of 
other permutation orbifolds in which the orbifold group is smaller than the full symmetric group, but still large enough to admit a holographic dual,  see \eg \cite{Hartman:2014oaa,Haehl:2014yla,Belin:2014fna,Belin:2015hwa}. In particular, the permutation orbifolds $(\mathbb{T}^2)^{MN}/(S_{M}\times S_N)$ are directly related to the large level limit of extended Kazama-Suzuki models, which in turn should be dual to suitable higher spin theories on AdS$_3$ \cite{Creutzig:2013tja,Candu:2014yva}. It would therefore be very interesting to understand whether there is a bulk string theory that is dual to $(\mathbb{T}^2)^{MN}/(S_{M}\times S_N)$.

\section*{Acknowledgements}
We would like to thank Ofer Aharony, Micha Berkooz, Justin David, Kevin Ferreira, Jerome Gauntlett and Ida Zadeh for discussions. 
The work of SD is supported by the NCCR SwissMAP, funded by the Swiss National Science Foundation. We gratefully acknowledge the hospitality of the Galileo Galilei Institute for Theoretical Physics (GGI), Florence and INFN for partial financial support during the programme `New developments in AdS$_3$/CFT$_2$ holography'.
 
\appendix

\section{Dihedral groups} \label{app:dihedral groups}
Throughout the paper we shall use the following presentation of the dihedral groups:
\be 
\mathrm{D}_n =\left\langle \, R,\, S\, \left|\, R^n=S^2=(RS)^2=1\, \right. \right\rangle\ .
\ee
We will frequently refer to $R$ as rotation and to $S$ as reflection. 
\subsection*{Representations} \label{app:representations}
The finite-dimensional representations of $\mathrm{D}_n$ are well known. For every $n$, there are two one-dimensional irreducible representation, which we denote by $\rho_+$ and $\rho_-$ satisfying
\be 
\rho_\pm (R)=1\ , \quad \rho_\pm (S)=\pm 1\ . \label{rho pm}
\ee
For even $n$, there are two further one-dimensional irreducible representations $\widetilde{\rho}_+$ and $\widetilde{\rho}_-$ satisfying
\be 
\widetilde{\rho}_\pm (R)=-1\ , \quad \widetilde{\rho}_\pm (S)=\pm 1\ . \label{widetilde rho pm}
\ee
Furthermore there are two-dimensional irreducible representations $\rho_j$, $j=1$, 2, \dots, $\lfloor \tfrac{n-1}{2} \rfloor$ satisfying
\be 
\rho_j(R)=\begin{pmatrix} \mathrm{e}^{\frac{2\pi\mathrm{i} j}{n}} & 0 \\ 0 & \mathrm{e}^{-\frac{2\pi\mathrm{i} j}{n}}\end{pmatrix}\ , \quad \rho_j(S)=\begin{pmatrix}0 & (-1)^j \\ (-1)^j & 0\end{pmatrix}\ . \label{rho i}
\ee
The $(-1)^j$ can be removed by a change of basis, but it will be convenient to retain these signs. Throughout the paper we shall identify $\rho_0 \cong \rho_+ \oplus \rho_-$ and  $\rho_{n/2} \cong \widetilde{\rho}_+ \oplus \widetilde{\rho}_-$ for $n$ even. One can easily confirm that this is formally true. Although $\rho_{n/2}$ is then not irreducible, this simplifies many formulas.

\subsection*{Tensor products}
We shall need some tensor products of the irreducible representations of $\mathrm{D}_n$, in particular
\begin{align}
\rho_{(-1)^\epsilon} \otimes \rho_{(-1)^\eta} &\cong \rho_{(-1)^{\epsilon+\eta}}\ , \\
\rho_{(-1)^\epsilon} \otimes \rho_i &\cong \rho_i\ , \\
\rho_i \otimes \rho_j &\cong \rho_{\min(i+j,n-i-j)} \oplus \rho_{\abs{i-j}}\ ,
\end{align}
Here $\epsilon,\eta \in \{0,1\}$.

\subsection*{Branching rules $\mathrm{SO}(4) \to \mathrm{D}_n$} \label{app:branching rules}
Since $\mathrm{D}_n $ is embedded into $\mathrm{SO}(4)$, the branching rules $\mathrm{SO}(4) \to \mathrm{D}_n$ will be important for us. The embedding is defined by twice the fundamental representation. By an appropriate change of basis, we can achieve that $\rho_1$ maps actually to $\mathrm{O}(2)$, so that $2\rho_1$ maps to $\mathrm{SO}(4)$ in this basis. Thus the defining branching rule is $(\mathbf{2},\mathbf{2}) \mapsto 2\rho_1$. In order to treat also spinor representations, we lift this embedding to its double cover, \ie a double cover of $\mathrm{D}_n$ is embedded into $\mathrm{SU}(2) \times \mathrm{SU}(2)$. For this, we embed $R$ into the Cartan torus of the first $\mathrm{SU}(2)$
\be 
R \mapsto \left(\begin{pmatrix} \mathrm{e}^{\frac{2\pi \mathrm{i}}{n}} & 0 \\ 0 & \mathrm{e}^{-\frac{2\pi \mathrm{i}}{n}} \end{pmatrix} ,\ \mathds{1}\right)\ .
\ee
$S$ is then embedded as 
\be 
S \mapsto \left(\begin{pmatrix}0 & 1\\ -1 & 0\end{pmatrix},\ \begin{pmatrix}0 & 1\\ -1 & 0\end{pmatrix}\right) \ .
\ee
Then one can check that under the double cover map $\mathrm{SU}(2) \times \mathrm{SU}(2) \to \mathrm{SO}(4)$, this embedding yields indeed the embedding of $\mathrm{D}_n$ in $\mathrm{SO}(4)$ given above. However, we see that in $\mathrm{SU}(2) \times \mathrm{SU}(2)$, we have $S^2=-1$, so this really defines an embedding of a double cover or equivalently a projective representation. 

In the main text we need the branchings of $(\mathbf{2},\mathbf{1})$ and $(\mathbf{1},\mathbf{2})$ to $\mathrm{D}_n$. These branchings give a priori only projective representations of $\mathrm{D}_n$. We will however see that they can be lifted to actual representations.

Regarding the representation $(\mathbf{2},\mathbf{1})$, the representations of $R$ and $S$ are 
\be 
\rho(R)=\begin{pmatrix} \mathrm{e}^{\frac{2\pi \mathrm{i}}{n}} & 0 \\ 0 & \mathrm{e}^{-\frac{2\pi \mathrm{i}}{n}} \end{pmatrix}
\ , \quad \rho(S)=\begin{pmatrix}
0 & 1 \\ -1 & 0
\end{pmatrix}\ .
\ee
As we have noticed before, $\rho(S)$ does not square to one, so this defines only a projective representation with cocycle
\be 
c(R^aS^\epsilon,R^b S^\eta)=(-1)^{\epsilon\eta}\ .
\ee
However, this cocycle is trivial since we can write
\be 
c(R^aS^\epsilon,R^b S^\eta)=f(R^aS^\epsilon R^b S^\eta)f(R^a S^\epsilon)^{-1}f(R^b S^\eta)^{-1}
\ee
with 
\be 
f(R^aS^\epsilon)=\mathrm{e}^{\frac{\pi\mathrm{i}}{2} \epsilon^2}\ ,
\ee
which just corresponds to multiplying $\rho(S)$ with $\mathrm{i}$. Thus, the projective representation is projectively equivalent to the representation $\rho_1$ and we conclude that the branching rule equals
\be 
(\mathbf{2},\mathbf{1}) \mapsto \rho_1\ .
\ee
The story is similar for $(\mathbf{1},\mathbf{2})$, the only difference being that the representation of $R$ is trivial. Thus, in this case the branching rule is
\be 
(\mathbf{1},\mathbf{2}) \mapsto \rho_+\oplus \rho_-\ .
\ee

\subsection{The fundamental representation over the integers}\label{app:integers}
In this subsection, we provide some details about what happens when one considers the representations over the integers. We shall mainly concentrate on the defining representation $\rho_1$. As was mentioned in Section~\ref{sec:dihedral}, this representation only exists over the integers for $n=1$, $2$, $3$, $4$ or $6$. In addition, there are two different representations over the integers for $n=1$, $2$ and $3$, which we denote by type (1) and (2), respectively. In this appendix we give explicit matrix realisations for these representations. 
\medskip 

\begin{align}
\mathrm{D}_1^{(1)}&=\left\langle R=\mathds{1}_2\ , \quad S=\begin{pmatrix}
-1 & 0 \\ 0 & 1
\end{pmatrix} \right\rangle\ , \\
\mathrm{D}_1^{(2)}&=\left\langle R=\mathds{1}_2\ , \quad S=\begin{pmatrix}
0 & 1 \\ 1 & 0
\end{pmatrix} \right\rangle\ , \\
\mathrm{D}_2^{(1)}&=\left\langle R=\begin{pmatrix}
-1 & 0 \\ 0 & -1
\end{pmatrix} , \quad S=\begin{pmatrix}
-1 & 0 \\ 0 & 1
\end{pmatrix} \right\rangle\ , \\
\mathrm{D}_2^{(2)}&=\left\langle R=\begin{pmatrix}
-1 & 0 \\ 0 & -1
\end{pmatrix} , \quad S=\begin{pmatrix}
0 & 1 \\ 1 & 0
\end{pmatrix} \right\rangle\ , \\
\mathrm{D}_3^{(1)}&=\left\langle R=\begin{pmatrix}
0 & -1 \\ 1 & -1
\end{pmatrix} , \quad S=\begin{pmatrix}
1 & 0 \\ 1 & -1
\end{pmatrix} \right\rangle\ , \\
\mathrm{D}_3^{(2)}&=\left\langle R=\begin{pmatrix}
0 & -1 \\ 1 & -1
\end{pmatrix} , \quad S=\begin{pmatrix}
1 & -1 \\ 0 & -1
\end{pmatrix} \right\rangle\ .
\end{align}
For $\mathrm{D}_3$, the three $\mathrm{D}_1$ subgroups, we considered in Section~\ref{sec:CFT} are generated by $S$, $RS$ and $R^2S$, respectively. 
We therefore have that 
\begin{align}
\mathrm{D}_3^{(1)}\ &:\quad S=\begin{pmatrix}
1 & 0 \\ 1 & -1
\end{pmatrix}\ , \quad RS=\begin{pmatrix}
-1 & 1 \\ 0 & 1
\end{pmatrix}\ , \quad R^2S=\begin{pmatrix}
0 & -1 \\ -1 & 0
\end{pmatrix}\ , \\
\mathrm{D}_3^{(2)}\ &:\quad S=\begin{pmatrix}
1 & -1 \\ 0 & -1
\end{pmatrix}\ , \quad RS=\begin{pmatrix}
0 & 1 \\ 1 & 0
\end{pmatrix}\ , \quad R^2S=\begin{pmatrix}
-1 & 0 \\ -1 & 1
\end{pmatrix}\ .
\end{align}
It is simple to check that all of these six matrices are conjugate to $\begin{pmatrix}
0 & 1 \\ 1 & 0
\end{pmatrix}$ over the integers. Hence all of these six $\mathrm{D}_1$ subgroups are of type (2). 
\smallskip

$\mathrm{D}_4$ is represented by the matrices
\be 
\mathrm{D}_4=\left\langle R=\begin{pmatrix}
0 & -1 \\ 1 & 0
\end{pmatrix} , \quad S=\begin{pmatrix}
-1 & 0 \\ 0 & 1
\end{pmatrix} \right\rangle\ .
\ee
The two $\mathrm{D}_2$ subgroups are generated by $\langle R^2, S \rangle$ and $\langle R^2, RS \rangle$. To determine their type, we compute
\be 
S=\begin{pmatrix}
-1 & 0 \\ 0 & 1
\end{pmatrix}\ , \quad RS=\begin{pmatrix}
0 & -1 \\ -1 & 0
\end{pmatrix}\ .
\ee
It is again simple to see that the first matrix is of type (1), while the second is of type (2).

Finally, $\mathrm{D}_6$ is represented by
\be 
\mathrm{D}_6=\left\langle R=\begin{pmatrix}
1 & -1 \\ 1 & 0
\end{pmatrix} , \quad S=\begin{pmatrix}
1 & -1 \\ 0 & -1
\end{pmatrix} \right\rangle\ .
\ee
The three $\mathrm{D}_2$-subgroups are generated by $\langle R^3, S \rangle$, $\langle R^3, RS\rangle$ and $\langle R^3, R^2 S \rangle$. We have
\be 
S=\begin{pmatrix}
1 & -1 \\ 0 & -1
\end{pmatrix}\ , \quad RS=\begin{pmatrix}
1 & 0 \\ 1 & -1
\end{pmatrix}\ , \quad R^2S=\begin{pmatrix}
0 & 1 \\ 1 & 0
\end{pmatrix}\ .
\ee
Again all of these matrices are of type (2), as claimed in the main text.

\section{$\mathcal{N}=4$ and $\mathcal{N}=2$ multiplets} \label{app:N4N2}
In this appendix, we explain how the $\mathcal{N}=4$ multiplets  decompose into $\mathcal{N}=2$ multiplets. In the following we shall work with the global subalgebra, since this is the relevant symmetry algebra of supergravity, \ie we are looking at the branching
\be 
\mathfrak{su}(1,1|2) \longrightarrow \mathfrak{osp}(1,1|2)\ .
\ee
Consider a short multiplet of $\mathfrak{su}(1,1|2)$. This has character
\be \label{N4char}
\chi^{\mathcal{N}=4}_{\ell}(z,\tau)=\frac{q^\ell}{1-q} \Bigl(\chi_\ell(z)+2q^{\frac{1}{2}} \chi_{\ell-\frac{1}{2}}(z)+q\chi_{\ell-1}(z) \Bigr)\ ,
\ee
where $\chi_\ell(z)$ denotes the spin $\ell$ $\mathfrak{su}(2)$ character. As always, $q=\mathrm{e}^{2\pi\mathrm{i}\tau}$ and $y=\mathrm{e}^{2\pi\mathrm{i}z}$. In contrast, an $\mathfrak{osp}(1,1|2)$ multiplet has character
\begin{align} 
\chi_{j,h}^{{\cal N}=2}(z,\tau)=\frac{q^h}{1-q}\begin{cases}
y^j+q^{\frac{1}{2}}y^{j-\frac{1}{2}} & \qquad j=h \\
y^j+q^{\frac{1}{2}}y^{j+\frac{1}{2}} & \qquad j=-h \\
y^j+q^{\frac{1}{2}}(y^{j-\frac{1}{2}}+y^{j+\frac{1}{2}})+q y^j & \qquad -h<j<h \ , 
\end{cases}
\end{align}
where the three cases correspond to chiral, anti-chiral and long representations, respectively. We shall also write $\chi^{\mathcal{N}=2}_{\ell,\ell}\equiv \chi^{\mathcal{N}=2}_\ell$ for chiral representations. Then it is easy to see that we have the decomposition
\be 
\chi^{\mathcal{N}=4}_\ell(z,\tau)=\sum_{j=-\ell}^\ell \chi_{j,\ell}^{\mathcal{N}=2}(z,\tau)\ ,
\ee
where the bottom and top components of this sum are short. 

In our setup, we break the $\mathcal{N}=4$ symmetry by having the different states in the multiplet transform in different representations of the dihedral group. (This breaks the $\mathcal{N}=4$ symmetry to $\mathcal{N}=2$.) Using the modified $\mathfrak{su}(2)$ characters 
\be 
\widetilde{\chi}_\ell(z)=\sum_{j=-\ell}^\ell \rho_{(-1)^{\ell-j}} \, y^j\ ,
\ee
the ${\cal N}=4$ character (\ref{N4char}) becomes 
\be \label{modchar}
\widetilde{\chi}^{\mathcal{N}=4}_{\ell}(z,\tau)=\frac{q^\ell}{1-q} \Bigl(\rho_+\widetilde{\chi}_\ell(z)+q^{\frac{1}{2}} (\rho_++\rho_-)\widetilde{\chi}_{\ell-\frac{1}{2}}(z)+\rho_-\widetilde{\chi}_{\ell-1}(z)\Bigr)\ .
\ee
This modified character still respects  the $\mathcal{N}=2$ subalgebra; indeed, we have 
\be 
\widetilde{\chi}^{\mathcal{N}=4}_\ell(z,\tau)=\sum_{j=-\ell}^\ell \rho_{(-1)^{\ell-j}}\, \chi_{j,\ell}^{\mathcal{N}=2}(z,\tau)\ .
\ee
Similarly, we can also describe the field content of a modified $\mathcal{N}=4$ multiplet in terms of modified $\mathfrak{su}(2)$ representations. Following \cite{deBoer:1998kjm}, we will continue to denote the short modified $\mathcal{N}=4$ multiplets by $(\mathbf{m})_\mathrm{S}$; it then consists of the fields 
\be
\begin{array}{lc}
h=h_0 \qquad &  \rho_+(\mathbf{m})  \\
h=h_0+\tfrac{1}{2} \qquad \qquad & (\rho_+ \oplus \rho_-)(\mathbf{m-1}) \\
h=h_0+1  \qquad &  \rho_- (\mathbf{m-2})\ ,
\end{array}
\ee
where $m=2h_0+1$.
This follows from the fact that one of the doublet of supercharges of the $\mathcal{N}=4$ algebra transforms in the representation $\rho_+$, the other in the $\rho_-$. The multiplets that are obtained by tensoring left- and right-movers will be denoted by 
 $(\mathbf{m},\mathbf{n})_\mathrm{S}$.

\section{Some properties of the elliptic genus of $(\mathbb{T}^2/\mathbb{Z}_2)\times (\mathbb{T}^2/\mathbb{Z}_2)$}\label{appC}

 Let us consider the elliptic genus of $(\mathbb{T}^2/\mathbb{Z}^2) \times (\mathbb{T}^2/\mathbb{Z}^2)$ and some properties of its coefficients. As before, we denote the doubled elliptic genus ($z\mapsto 2z$) by $\tZ(z,\tau)$. The explicit expressions for the different sectors are then 
 \begin{align}\label{d-eg}
\tZ_{\rm UU}(z,\tau) &= 4 \left[ \frac{\vt_2(2z|\tau)}{\vt_2(0|\tau)}\right]^2 \\
\tZ_{\rm TT}(z,\tau) &= 4 \left[ \frac{\vt_3(2z|\tau)}{\vt_3(0|\tau)} + \frac{\vt_4(2z|\tau)}{\vt_4(0|\tau)}\right]^2 \\
\tZ_{\rm UT}(z,\tau) &= 8  \frac{\vt_2(2z|\tau)}{\vt_2(0|\tau)}\left[ \frac{\vt_3(2z|\tau)}{\vt_3(0|\tau)} + \frac{\vt_4(2z|\tau)}{\vt_4(0|\tau)}\right] \ .
 \end{align}
We have denoted the coefficients of the Fourier expansion by $c_*(m,l)$, where `$*$' refers to the sectors UU, UT or TT. We shall also use the quasiperiodicity properties of the Jacobi theta functions, which implies that the coefficients of $\tZ_*$ satisfy
\begin{align}\label{pp}
c_*(m,\ell) = c_*(16m-\ell ^2, \ell\text{ mod }8) \ .
\end{align}
This leads to 
\begin{align}\label{id1}
\sum_{\ell \, \in \, \text{even},\, m} c_*(16m-\ell^2, \ell\text{ mod }8 )q^m  &= \tfrac{1}{2} \bigl(\tZ_*(0,\tau)+\tZ_*(\tfrac{1}{2},\tau)\bigr)\ , \\ 
\qquad \sum_{\ell \, \in \, \text{odd},\, m} c_*(16m-\ell^2, \ell\text{ mod }8 )q^m  &= \tfrac{1}{2}\bigl(\tZ_*(0,\tau)-\tZ_*(\tfrac{1}{2},\tau)\bigr) \ .\label{id2}
\end{align}
Owing to the additional dependence on $(\ell \text{ mod } 8)$, we shall require two other properties of the coefficients. In order to derive them, we consider the generating function
\begin{align}
\sum_{\ell \, \in \, \text{even},\, m} c_*(16m-\ell^2, \ell+4\text{ mod }8 )q^m  &= \sum_{\ell' \, \in \, \text{even},\, m} c_{*}(16m-16+8\ell'-\ell'^2, \ell'\text{ mod }8 )\, q^m \nn \\
&= \sum_{\ell' \, \in \, \text{even},\, m} c_*(m-1+\ell'/2,\ell')\, q^m \nn \\
&= \sum_{\ell' \, \in \, \text{even},\, m'} c_*(m',\ell')\, q^{m'+1} \, y^{\ell'} \bigg|_{z=-\tau/2} \nn \\
&= \half q \left[ \tZ_*(-\tfrac{\tau}{2},\tau) + \tZ_*(-\tfrac{\tau}{2}+\tfrac{1}{2},\tau)\right]  \ . \label{id3}
\end{align}
Here, in the first equality we have redefined the variable $\ell'=\ell+4$. The second equality uses \eqref{pp}. In the final steps we have cast the expression in a form so that it can be re-expressed in terms of the doubled elliptic genus.  
 Similarly, the above identity for summing over $\ell \, \in \, \text{odd}$ can be found to be 
\begin{align}
\sum_{l \, \in \, \text{odd},\, m} c_*(16m-\ell^2, \ell+4\text{ mod }8 )\, q^m  =\half q \left[ \tZ_*(-\tfrac{\tau}{2},\tau) - \tZ_*(-\tfrac{\tau}{2}+\tfrac{1}{2},\tau)\right]  \ .\label{id4}
\end{align}

The values of the elliptic genus at these points can be found using the quasi-periodicity properties of the theta functions. We list the results for the three sectors below. 
\begin{align}
\sum_{\ell \, \in \, \text{even},\, m}\!\!\!\!\! c_{\rm UU}(16m-\ell^2, \ell\text{ mod }8 )q^m  &= 4 \ , \quad
 \sum_{\ell\, \in \, \text{odd},\, m} \!\!\!\!\! c_{\rm UU}(16m-\ell^2, \ell\text{ mod }8 )q^m   = 0\ , \nn \\
\sum_{\ell\, \in \, \text{even},\, m}\!\!\!\!\! c_{\rm TT}(16m-\ell^2, \ell\text{ mod }8 )q^m  &= 16\ , \quad 
 \sum_{\ell \, \in \, \text{odd},\, m}\!\!\!\!\! c_{\rm TT}(16m-\ell^2,\ell\text{ mod }8 )q^m   = 0\ , \nn \\
\sum_{\ell \, \in \, \text{even},\, m}\!\!\!\!\! c_{\rm UT}(16m-\ell^2, \ell\text{ mod }8 )q^m  &= 0\ , \quad
 \sum_{\ell \, \in \, \text{odd},\, m}\!\!\!\!\! c_{\rm UT}(16m-\ell^2, \ell\text{ mod }8 )q^m   = 16\ ,\label{l mod 8}  \\
\sum_{\ell \, \in \, \text{even},\, m}\!\!\!\!\! c_{\rm UU}(16m-\ell^2, \ell+4\text{ mod }8 )q^m  &= 4 \ ,  \
 \sum_{\ell \, \in \, \text{odd},\, m}\!\!\!\!\! c_{\rm UU}(16m-\ell^2, \ell+4\text{ mod }8 )q^m   = 0\ , \nn \\
\sum_{\ell \, \in \, \text{even},\, m}\!\!\!\!\! c_{\rm TT}(16m-\ell^2, \ell+4\text{ mod }8 )q^m  &= 0\ , \  
 \sum_{\ell \, \in \, \text{odd},\, m}\!\!\!\!\! c_{\rm TT}(16m-\ell^2, \ell+4\text{ mod }8 )q^m   = 0 \ ,\nn \\
\sum_{\ell \, \in \, \text{even},\, m}\!\!\!\!\! c_{\rm UT}(16m-\ell^2, \ell+4\text{ mod }8 )q^m  &= 0\ , \ 
 \sum_{\ell \, \in \, \text{odd},\, m}\!\!\!\!\! c_{\rm UT}(16m-\ell^2, \ell+4\text{ mod }8 )q^m   = 0\ .\label{l+4 mod 8}
\end{align}
The above analysis can also be directly generalized to the other cases, \ie for the general form of the elliptic genus of eq.~\eqref{doubled elliptic genus}.

\end{document}